%% file: hs.tex
\documentclass[apjl]{emulateapj}
\usepackage{comment}
\usepackage{ifthen}
\usepackage{lineno}
\usepackage{lipsum}

\input{newcommand_gen.tex}

\input{newcommand_spec.tex}

\newboolean{emulateapj}
\setboolean{emulateapj}{true}

\newboolean{astroph}
\setboolean{astroph}{true}

\shortauthors{Bakos et al.}
\shorttitle{The HATSouth network}
\ifthenelse{\boolean{emulateapj}}{
    \newcommand{\titledag}{$\dagger$}
}{
    \newcommand{\titledag}{\dagger}
}

\ifthenelse{\boolean{emulateapj}}{
    \newcommand{\titlestar}{$\star$}
}{
    \newcommand{\titlestar}{\star}
}

\begin{document}


\title{HATSouth: a global network of fully automated identical
	wide-field telescopes
\altaffilmark{\titledag}}

\author{
        G.~\'A.~Bakos\altaffilmark{1,2,\titlestar},
		Z.~Csubry\altaffilmark{1,2},
		K.~Penev\altaffilmark{1,2},
        D.~Bayliss\altaffilmark{3},
		A.~Jord\'an\altaffilmark{4},
        C.~Afonso\altaffilmark{5},
        J.~D.~Hartman\altaffilmark{1,2},
        T.~Henning\altaffilmark{5},
		G.~Kov\'acs\altaffilmark{6},
        R.~W.~Noyes\altaffilmark{2},
        B.~B\'eky\altaffilmark{2},
 		V.~Suc\altaffilmark{4},
        B.~Cs\'ak\altaffilmark{5},
		M.~Rabus\altaffilmark{4},
        J.~L\'az\'ar\altaffilmark{7},
        I.~Papp\altaffilmark{7},
        P.~S\'ari\altaffilmark{7},
		P.~Conroy\altaffilmark{3},
		G.~Zhou\altaffilmark{3},
        P.~D.~Sackett\altaffilmark{3},	
		B.~Schmidt\altaffilmark{3},
		L.~Mancini\altaffilmark{5},
		D.~D.~Sasselov\altaffilmark{2},
        K.~Ueltzhoeffer\altaffilmark{8}
}

\altaffiltext{1}{Department of Astrophysical Sciences,
	Princeton University, NJ 08544, USA;
	email: gbakos@astro.princeton.edu}

\altaffiltext{2}{Harvard-Smithsonian Center for Astrophysics,
	Cambridge, MA, USA}

\altaffiltext{$\star$}{Alfred P.~Sloan Research Fellow}

\altaffiltext{3}{The Australian National University, Canberra,
	Australia}

\altaffiltext{4}{Departamento de Astronom\'ia y Astrof\'isica, Pontificia, 
	Universidad Cat\'olica de Chile, 7820436 Macul, Santiago, Chile}

\altaffiltext{5}{Max Planck Institute for Astronomy,
	Koenigstuhl 17, 69120 Heidelberg, Germany}

\altaffiltext{6}{Institute of Astronomy, University of Cambridge, 
	Cambridge, UK}

\altaffiltext{7}{Hungarian Astronomical Association, Budapest,
	Hungary}

\altaffiltext{8}{Goethe University Frankfurt, 
	Bernstein Center for Computational Neuroscience Heidelberg/Mannheim}

\altaffiltext{$\dagger$}{
The HATSouth hardware was acquired by NSF MRI NSF/AST-0723074, and is
owned by Princeton University.  The HATSouth network is operated by a
collaboration consisting of Princeton University (PU), the Max Planck
Institute for Astronomy (MPIA), and the Australian National University
(ANU).  The station at Las Campanas Observatory (LCO) of the Carnegie
Institution for Science, is operated by PU in conjunction with
collaborators at the Pontificia Universidad Cat\'olica de Chile (PUC),
the station at the High Energy Spectroscopic Survey (HESS) site is
operated in conjunction with MPIA, and the station at Siding Springs
Observatory (SSO) is operated jointly with ANU.
}

\begin{abstract}
\setcounter{footnote}{6}
HATSouth is the world's first network of automated and homogeneous
telescopes that is capable of year-round 24-hour monitoring of
positions over an entire hemisphere of the sky.  The primary scientific
goal of the network is to discover and characterize a large number of
transiting extrasolar planets, reaching out to long periods and down to
small planetary radii.
HATSouth achieves this by monitoring extended areas on the sky,
deriving high precision \lcs\ for a large number of stars, searching
for the signature of planetary transits, and confirming planetary
candidates with larger telescopes.
HATSouth employs six telescope units spread over three prime locations
with large longitude separation in the southern hemisphere (Las
Campanas Observatory, Chile; HESS site, Namibia; Siding Spring
Observatory, Australia).  Each of the HATSouth units holds four 0.18\,m
diameter f/2.8 focal ratio telescope tubes on a common mount producing
an $8.2\arcdeg\times8.2\arcdeg$ field-of-view on the sky, imaged using
four \ccdsize{4K} CCD cameras and Sloan $r$ filters, to give a pixel
scale of 3.7\pxs.  The HATSouth network is capable of continuously
monitoring 128 square arc-degrees at celestial positions moderately
close to the anti-solar direction.  We present the technical details of
the network, summarize operations, and present detailed weather
statistics for the three sites.  Robust operations have meant that on
average each of the six HATSouth units has conducted observations on
$\sim 500$ nights over a two-year time period, yielding a total of more
than 1 million science frames at four minute integration time, and
observing $\sim10.65$\,hours per day on average.  We describe the
scheme of our data transfer and reduction from raw pixel images to
trend-filtered \lcs\ and transiting planet candidates.  Photometric
precision reaches $\sim 6$\,mmag at 4 minute cadence for the brightest
non-saturated stars at $r\approx10.5$. We present detailed transit
recovery simulations to determine the expected yield of transiting
planets from HATSouth.  We highlight the advantages of networked
operations, namely, a threefold increase in the expected number of
detected planets, as compared to all telescopes operating from the same
site.
\setcounter{footnote}{0}
\end{abstract}



\section{Introduction}
\label{sec:introduction}

Robotic telescopes first appeared about 40 years ago.  The primary
motivations for their development included cost efficiency, achieving
consistently good data quality, and diverting valuable human time from
monotonous operation into research.  The first automated and
computer-controlled telescope was the 0.2\,m reflector of Washburn
Observatory \citep{mcnall:1968}.  Another noteworthy development was
the Automated Photometric Telescope \citep[APT;][]{boyd:1984} project,
which achieved a level of automation that enabled more than two decades
of unmanned operations.  As computer technology, microelectronics,
software, programming languages, and interconnectivity (Internet) have
developed, remotely-operated or fully-automated (often referred to as
autonomous) telescopes have become widespread \citep[see][for a
review]{ct:2010}.  A few prime examples are: the 0.75\,m Katzman
Automatic Imaging Telescope \citep[KAIT;][]{filippenko:2001} finding a
large number of supernovae; the Robotic Optical Transient Search
Experiment-I (ROTSE-I) instrument containing four 0.11\,m diameter
lenses, which for exampled detected the spectacular $V=8.9$\,mag
optical afterglow of a gamma ray burst at redshift of $z\approx1$
\citep{akerlof:1999}; the LIncoln Near Earth Asteroid Research
\citep[LINEAR;][]{stokes:1998} and Near Earth Asteroid Tracking
\cite[NEAT;][]{pravdo:1999} projects using 1\,m-class telescopes and
discovering over a hundred thousand asteroids to date; the All Sky
Automated Survey \citep[ASAS;][]{pojmanski:2002} employing a 0.1\,m
telescope to scan the entire sky and discover $\sim 50000$ new
variables; the Palomar Transient Factory \citep[PTF;][]{rau:2009}
exploring the optical transient sky, finding on average one transient
every 20 minutes, and discovering $\sim 1500$ supernovae so far; the
Super Wide Angle Search for Planets \citep[SuperWASP;][]{pollacco:2006}
and Hungarian-made Automated Telescope Network
\citep[HATNet;][]{bakos:2004} projects employing 0.1\,m telescopes and
altogether discovering $\gtrsim 100$ transiting extrasolar planets.

To improve the phase coverage of time-variable phenomena, networks of
telescopes distributed in longitude were developed.  We give a few
examples below.  One such early effort was the Smithsonian
Astrophysical Observatory's satellite tracker project
\citep{whipple:1956,henize:1958}, using almost identical hardware
(Baker-Nunn cameras) at 12 stations around the globe, including
Cura\c{c}ao and Ethiopia.  This network was manually operated.  Another
example is the Global Oscillation Network Group project
\citep[GONG;][]{harvey:1988}, providing Doppler oscillation
measurements for the Sun, using 6 stations with excellent phase
coverage for solar observations ($|\delta|\!<23.5\arcdeg$). The Whole
Earth Telescope \citep[WET;][]{nather:1990} uses existing (but quite
inhomogeneous) 1\,m-class telescopes at multiple locations in organized
campaigns to monitor variable phenomena \citep{provencal:2012}.  The
PLANET collaboration \citep{albrow:1998} employed existing 1\,m-class
telescopes to establish a round-the-world network, leading to the
discovery of several planets via microlensing anomalies.  Similarly,
RoboNet \citep{tsapras:2009} used 2\,m telescopes at Hawaii, Australia,
and La Palma to run a fully automated network to detect planets via
microlensing anomalies.  ROTSE-III \citep{akerlof:2003} has been
operating an automated network of 0.5\,m telescopes for the detection
of optical transients, with stations in Australia, Namibia, Turkey and
the USA.

The study of transiting extrasolar planets (TEPs) has greatly benefited
from the development of automated telescopes and networks. 
\citet{mayor:2009} and \citet{howard:2010,howard:2011} concluded that
$\sim 1.7\%$ of dwarf stars harbor planets with radii between
$3\rearth$ and $32\rearth$ and periods less than $20$\,d; such planets
could be be detected by ground-based surveys such as ours.\footnote{The
choice of these limits is somewhat arbitrary, but does not change the
overall conclusions.  See \refsec{perf} for more details.} When coupled
with the geometric probability that these planets transit their host
stars as seen from the Earth, only $\sim 0.11\%$ of dwarf stars have
TEPs with the above parameters.  Further, in a brightness
limited sample with e.g.~$r\!<\!12$\,mag, only $\sim 40\%$ of the stars
are A5 to M5 dwarfs (enabling spectroscopic confirmation and planetary
mass measurement), thus fewer than 1 in 2000 of the $r\!<\!12$\,mag
stars will have a moderately ($>3\,\rearth$) large radius and short
period ($P<20$\,d) TEP.  Consequently, monitoring of tens
of thousands of stars at high duty cycle and homogeneously optimal data
quality is required for achieving a reasonable TEP detection yield.

\begin{figure*}[!ht]
\plotone{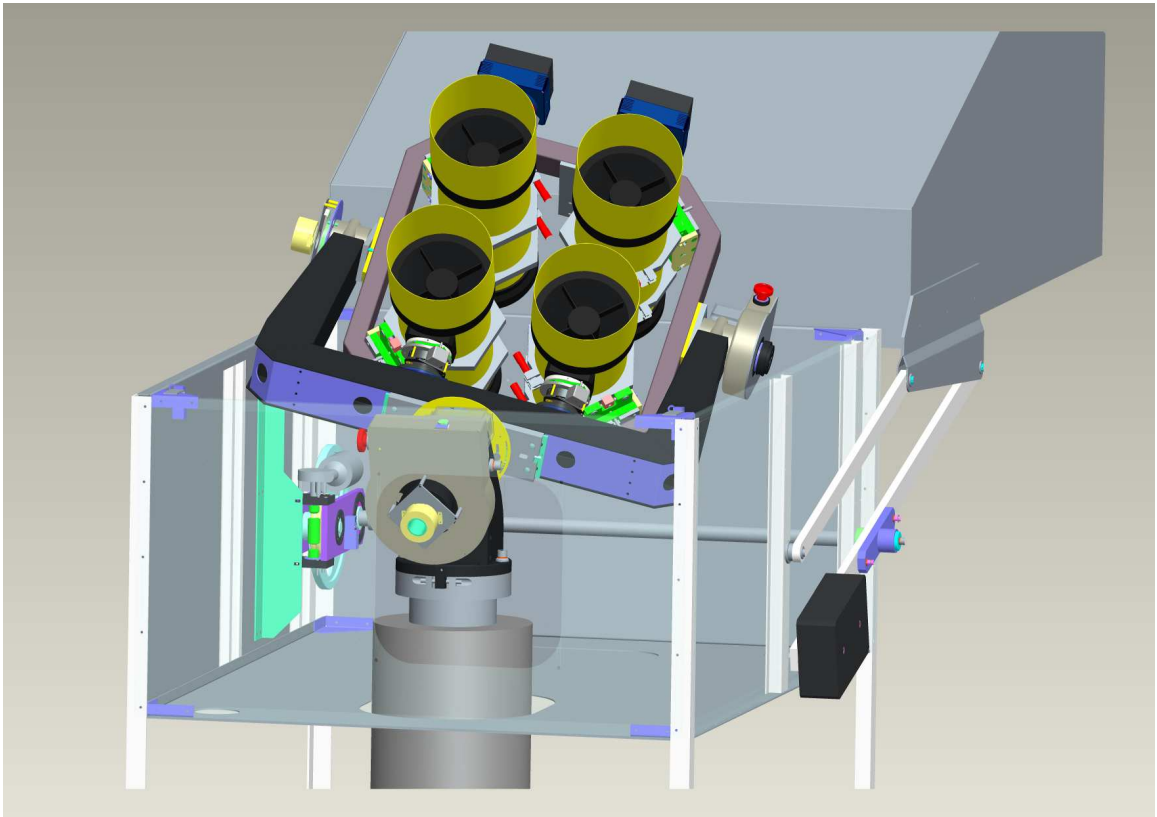}
\caption{
	Engineering model of the \hsfour\ unit, depicting the dome,
	telescope mount, optical tubes and CCDs.  The asymmetric clamshell
	dome can open/close with the telescope in any position.  The
	equatorial fork mount holds a large frame that supports the four
	astrographs and CCDs, tilted $\sim 4\arcdeg$ with respect to each
	other, and therefore capable of producing a mosaic image of
	$8\arcdeg\times8\arcdeg$.
\label{fig:eng}}
\end{figure*}

To date approximately 140 TEPs have been confirmed, characterized
with RVs to measure the planetary mass, and published.\footnote{
	See http://exoplanets.org \citep{wright:2011} for the list of
	published planets, and www.exoplanet.eu \citep{schneider:2011} for
	a compilation including unpublished results.  In this discussion we
	refer to the published planets, focusing only on those for which
	the RV variation of the star due to the planet has been measured.
}
These have been found primarily by photometric transit surveys
employing automated telescopes (and networks in several cases) such as
WASP \citep{pollacco:2006}, HATNet \citep{bakos:2004}, CoRoT
\citep{baglin:2006}, OGLE \citep{udalski:2002}, {\em Kepler}
\citep{borucki:2010}, XO \citep{mccullough:2005}, and TrES
\citep{alonso:2004}.  In addition, {\em Kepler} has found over 2000
strong planetary candidates, which have been instrumental in
determining the distribution of planetary radii.  Many ($\sim 40$) of
these planetary systems have been confirmed or ``validated''
\citep[][and references therein]{batalha:2012}, although not
necessarily by radial velocity measurements.  While the sample of $\sim
140$ fully confirmed planets with accurate mass measurements is large
enough to reveal tantalizing correlations among various planetary
(mass, radius, equilibrium temperature, etc.) and stellar (metallicity,
age) properties, given the apparent diversity of planets, it is still
insufficient to provide a deep understanding of planetary systems.  For
only the brightest systems is it currently possible to study extrasolar
planetary atmospheres via emission or transmission spectroscopy; the
faintest system for which a successful atmosphere study has been
performed is WASP-12, which has $V \approx 11.6$\,mag;
\citep{madhusudhan:2011}.  Similarly, it is only for the brightest
systems that one can obtain a high S/N spectrum in an exposure time
short enough to resolve the Rossiter-McLaughlin effect
\citep{holt:1893,schlesinger:1910,rossiter:1924,mclaughlin:1924}, and
thereby measure the projected angle between the planetary orbital axis
and the stellar spin axis.

The existing sample of ground-based detections of TEPs around bright
stars is highly biased toward Jupiter-size planets with periods shorter
than 5 days.  Only 13 of the $\sim 140$ RV-confirmed TEPs have masses
below $0.1\,\mjup$, and only 12 have periods longer than 10 days.  The
bias towards short periods is due not only to the higher geometric
probability of short-period transits, and relative ease of their
confirmation with spectroscopic (radial-velocity) observations, but
also to the low duty cycle of single-longitude surveys.  Although the
transiting hot Jupiters provide an opportunity to study the properties
of planets in an extreme environment, they are not representative of
the vast majority of planetary-mass objects in the Universe, which are
likely to be of lower mass, and on longer period orbits.  While other
planet-detection methods, such as microlensing, have proven to be
efficient at discovering long-period and low-mass planets
\citep{gould:2010,dong:2009}, these methods are primarily useful for
studying the statistical distributions of periods and masses of
planets, and cannot be used to study the other physical properties of
individual planets, which can only be done for TEPs.

In this paper we descript HATSouth, a set of new ground-based
telescopes which form a global and automated network with exactly
identical hardware at each site, and with true 24-hour coverage all
year around (for any celestial object in the southern hemisphere, and
``away'' from the Sun in a given season).  HATSouth is the first such
network, although many more are planned.  The Las Cumbres Observatory
Global Telescope \citep[LCOGT;][]{brown:2010}, SOLARIS
\citep{konacki:2011}, and the KMTNet \citep{kim:2010} will all form
global, homogeneous and automated networks when they are completed.

The HATSouth survey, in operation since late 2009, has the northern
hemisphere HATNet survey \citep{bakos:2004} as its heritage.  HATSouth,
however, has two important distinctions from HATNet, and from all other
ground-based transit surveys.  The first and most important is its
complete longitudinal coverage.  The network consists of six robotic
instruments distributed across three sites on three continents in the
southern hemisphere: Las Campanas Observatory (LCO) in Chile, the High
Energy Stereoscopic System (HESS) site in Namibia, and Siding Springs
Observatory (SSO) in Australia.  The geographical coordinates of these
sites are given in \reftab{specs} below.  The longitude distribution of
these observatories enables round-the-clock monitoring of selected
fields on the sky.  This greatly increases the detectability of TEPs,
particularly those with periods in excess of a few days.  This gives
HATSouth an order of magnitude higher sensitivity than HATNet to
planets with periods longer than 10 days, and its sensitivity towards
$P\approx15-20$\,d planets is better than HATNet's sensitivity at
$P\approx8$\,d .  This is encouraging given that HATNet has
demonstrated sensitivity in this regime with the discoveries of
HAT-P-15b \citep{kovacs:2010} and HAT-P-17b \citep{howard:2012} at
$P>10$\,d.  Note that for mid- to late-M dwarf parent stars, planets
with $\sim15$\,d periods lie in the habitable zone.

The second difference between HATSouth and HATNet is that each HATSouth
astrograph has a larger aperture than a HATNet telephoto lens (0.18\,m
vs.~0.11\,m), plus a slower focal ratio and lower sky background (per
pixel, and under the point spread function of a star), which allows
HATSouth to monitor fainter stars than HATNet.  Compared to
HATNet, this increases the overall number of dwarf stars observed at
1\% photometry over a year by a factor of $\sim 3$; more specifically
the number of K and M dwarf stars monitored effectively is increased by
factors of 3.1 and 3.6, respectively (the numbers take into account the
much larger surface density of dwarf stars and the somewhat smaller
field-of-fiew of HATSouth, along with slight differences in the
observing tactics).  This increases the expected yield of small-size
planets, and opens up the possibility of reaching to
the super-Earth range.  Furthermore, the ratio of dwarf stars to giant
stars that are monitored at 1\% photometric precision in the HATSouth
sample at $|b|\approx20\arcdeg$ is about twice\footnote{This ratio is
much higher closer to the galactic plane, and is close to unity at the
galactic pole.} that of HATNet, yielding a lower false alarm rate. 
Furthermore, despite greater stellar number densities, stellar crowding
is less than with HATNet due to HATSouth's three times finer spatial
(linear) resolution.  Note that while the stellar population monitored
is generally fainter than that of HATNet, they are still within the
reach of follow-up facilities.

The layout of the paper is as follows. In \refsecl{hw} we describe the
HATSouth hardware in detail, including the telescope units
(\refsec{hs4}), weather sensing devices (\refsec{wth}), and the
computer systems (\refsec{compsys}).  In \refsecl{csw} we detail the
instrument control software.  The HATSouth sites and operations are
laid out in \refsecl{sitop}.  We give details on the site specifics
(\refsec{sites}), the scheme of nightly operations (\refsec{oper}), and
present observing statistics for two years (\refsec{stat}).  Data flow
and analysis are described in \refsecl{dr}, and the expected planet
yield is calculated using detailed simulations in \refsecl{perf}.

\section{The HATSouth hardware}
\label{sec:hw}

Each of the three HATSouth stations hosts two fully automated
``observatories'', referred to as \hsfour\ units.  One \hsfour\ unit
holds 4 telescope tubes and 4 CCDs attached to a robotic mount, and
enclosed by a robotic dome.  The \hsfour\ units are controlled via
computers from a dedicated control building that is $\sim 10$\,meters
north of the telescopes.  The control building has multiple uses.  It
hosts a low-light web camera for monitoring telescopes (\refsec{wth}),
all the computing equipment (\refsec{compsys}), and all tools and spare
components used for telescope maintenance and repair.  The roof of the
building is populated by weather sensing devices and an all-sky camera
(\refsec{wth}).  In the following subsections we describe the hardware
components in more detail.

\subsection{The \hsfour\ unit}
\label{sec:hs4}

An \hsfour\ unit is a fully automated observatory consisting of the
following components:
\begin{itemize}
\item
A Fornax F150 equatorial fork mount,
\item 
Four 18\,cm aperture f/2.8 Takahashi hyperbolic astrograph
optical tube assemblies (OTAs),
\item
Four custom-built automated focuser units, each mounted to an OTA,
\item
Four Apogee U16m \ccdsize{4K} CCD detectors, each mounted to a focuser
unit,
\item
An asymmetric clamshell dome with weather sensing devices,
\item A weatherproof electronic box attached to the dome, with power
supplies, instrument control electronics and communication related
electronics,
\item
An instrument control and data acquisition computer that is responsible 
for controlling all
the above, and which is hosted in the nearby control building.
\end{itemize}

The four optical tubes are tilted with respect to each other to have a
small overlap along the edges of the individual fields of view (FOV),
and to produce a mosaic FOV spanning $8.2\arcdeg \times 8.2\arcdeg$ on
\ccdsize{8K} pixels altogether with a scale of $3.7\pxs$.  Since two
\hsfour\ instruments are located at each site, and they are pointed at
different fields, one site monitors a 128\sqarcdeg\ sky area.  Each of
the three HATSouth sites have the same field setup, and because of the
near-optimal longitude separation of the three sites, the HATSouth
network is capable of {\em continuously}\/ monitoring a 128\sqarcdeg\
sky area (in the anti-solar direction).  In the following we present
more information on the instrument parameters.  These are also
summarized in \reftab{specs}.  The engineering design is shown in
\reffig{eng}.

\begin{figure}[!ht]
\plotone{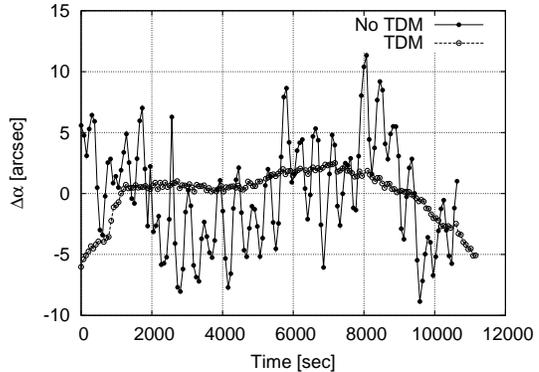}
\caption{
	Periodic errors (displacement in RA) of the HATSouth mount.  The
	solid curve with solid circles shows the performance of the mount
	without tracking correction during a series of 30\,s exposures. 
	Periodic errors with $\sim 10\arcsec$ peak-to-peak amplitude are
	seen.  The dashed curve with open circles shows the same
	experiment, repeated two days later, with the TDM tracking
	correction turned on.  Long term drifts have been removed by a
	simple fit of a second order polynomial, yet some trends remain (no
	astrometry and subsequent corrections were applied in these tests). 
	Using the TDM, the amplitude of the short term periodicity is
	reduced to $\sim 0.5\arcsec$ peak-to-peak amplitude where no drift
	is present.  Overall, including longer term drifts, this translates
	to just $\sim 1.2\arcsec$ (median) displacement over 4 minute
	integrations.
\label{fig:periodic}}
\end{figure}

\subsubsection{The telescope mount}
\label{sec:mount}

The Fornax F150 equatorial fork mount was designed, developed and
constructed by our team specifically for the HATSouth project.  Initial
polar alignment of the mount is performed using a polar telescope that
fits in the right ascension (RA) axis of the mount.  The alignment is
then refined by taking polar exposures and by adjusting the angle of
the polar axis using fine alignment screws.

The RA axis is fitted with three inductive proximity sensors; one for
the two end positions on the east and west, and the third one for the
so-called home position very close to the meridian.  Similarly, for the
declination (DEC) axis we have proximity sensors for the polar and
northern horizon end position, and one for the home position that
roughly coincides with the celestial equator.  When the mount reaches
either the end or home positions the relevant proximity sensor is
activated and alerts the electronics and the control software.  If any
of the end proximity sensors is activated during normal operation of
the mount, telescope motion is inhibited by the software.  If, for some
reason, the proximity sensors fail, we have a second level of
protection in the form of electronic limit switches just beyond the
proximity sensor positions.  If these are activated, then the motion of
the telescope mount in that direction is inhibited directly by
electronics, without relying on the control software.  Finally, if the
electronic limit switches fail, the mount hits the mechanical end
positions, and any further motion is taken up by a clutch system on
both the RA and DEC
axes.  
%
%
We found that this level of redundancy in safety measures is essential
for robust automated operation.

The exact position of the home sensors was measured during telescope
installation via astrometry, and is re-calibrated every time a change
in the hardware necessitates.  The hour angle of the RA home position
is determined to an accuracy of $\sim 25\arcsec$, and the declination
of the DEC home position to $\sim 10\arcsec$.  The mount has the
capability to find and settle on the home position from any starting
position within at most $\sim 200$\,seconds, and more typically in
1\,minute, using an iterative scheme and information from the proximity
sensors.  Following this homing procedure, which we perform at the
beginning of each night, and using the local sidereal time (based on
our GPS or the Network Time Protocol, see \refsec{compsys}), the
pointing of the mount is known at an accuracy of $\lesssim 1\arcmin$.

The RA gear of the mount consists of a 292\,mm diameter bronze cogwheel
with 192 teeth.  It is driven by a worm-drive, which, in turn, is
driven by a two-phase stepper motor through sprockets and gears (with
an additional 1:4.5 gear ratio).  The resulting overall gear ratio is
1:864, so one full turn of the stepper motor axis corresponds to a
1/864 turn of the RA axis.  The stepper motor receives the clock and
direction signals from the printer port of the control computer through
the control electronics.  One microstep of the motor corresponds to
0.5\arcsec\ resolution on the celestial equator (or 1/30 seconds of
time in RA), i.e.~the mount is driven at $\approx30.08$\,Hz during
sidereal rate tracking.  The mount has a massive fork with a span of
830\,mm holding a rectangular frame on its DEC axis.  This frame holds
the four Takahashi optical tube assemblies (OTAs) through a mechanism
that allows fine tilting of each optical tube in two perpendicular
directions, so as to achieve a well aligned mosaic image.  The main DEC
axis gear consists of a 195\,mm diameter cogwheel with 192 teeth.  It
is driven in a similar fashion to the RA gear, and one microstep on the
motor corresponds to 0.5\arcsec.  We set the maximal speed of the RA
and DEC axes to be 2.2\arcdeg/s (corresponding to 16\,KHz drive
frequency), and we typically ramp up in 50 steps over 3 degrees to the
maximal speed during coarse motion.  These parameters are fully
adjustable from our control software.

Our pointing accuracy (median offset from the desired position) using
this coarse motion is $\sim140\arcsec$ (RA) and $\sim55\arcsec$ (DEC)
on the sky without a telescope pointing model, but after correcting for
refraction.  The accuracy is primarily limited by various flexures in
the fork and the DEC frame, and by the imperfect polar alignment.  This
accuracy for coarse motion is adequate since it represents at most
0.5\% of our $8\arcdeg\times8\arcdeg$ FOV and we have the capability of
doing astrometry on our images at sub-arcsecond accuracy, if these
images contain at least a few hundred stars (typical numbers under
normal conditions are $\ordo(10^4)$).  Following coarse motion pointing
to a position, and initial astrometry, we then use fine motion to
correct the position of the telescope with a small angle movement at
low speed.  To correct for a significant backlash we added an encoder
to the DEC axis, and special electronics that together form a
closed-loop control of the position when performing fine motion
movements.  We also have an encoder on the RA axis, which is used for
precise sidereal rate tracking (see later).  Fine motion in RA is
performed by stopping or speeding up tracking.  After a few exposures,
we typically reach and maintain better than 10\arcsec\ accuracy (in
radial distance) while observing the same field.

Periodic tracking errors are inherently present in systems with
worm-and-wheel gearing.  It is common to have $\sim 10\arcsec$
positional variation in RA on the celestial equator.  We carried out
tracking tests to measure the periodic error of the Fornax F150 mounts. 
A selected field that culminates in zenith was observed for 2\,hours at
30-second cadence from hour angle $-1$\,hour to $+1$\,hour.  We found
periodic errors with a peak-to-peak amplitude of $\lesssim 10\arcsec$
(median), and measured the main period to be $\sim 450$\,seconds
(\reffig{periodic}).  This matches the period expected from the
engineering design.  The tests were performed without using tracking
correction (see below).

The movement of stars with respect to the CCD lattice leads to unwanted
noise in the photometry, and the above tracking error would correspond
to a large $\sim5.6\arcsec$ (1.5\,pixel) median displacement in RA
during our typical 240\,s integrations.  One common solution for
suppressing periodic errors is auto-guiding on stars by using separate
optics (or a pick-off mirror) and a guide CCD.  Autoguiding, however,
is a sensitive procedure that requires acquiring suitable guide-stars,
and it would lead to a large increase in hardware and operational
complexity that we were keen to avoid.  We thus used a hybrid solution,
developed by our team.  This is the ``Telescope Drive Master'' (TDM),
consisting of an encoder on the RA axis, and a closed-loop control
system that corrects the tracking clock signals based on the feedback
from the encoder \citep[see review by][]{diciccio:2011}.  Our tracking
precision with the TDM is greatly improved, as exhibited in
\reffig{periodic}.  The mean displacement during a 240\,s integration
is reduced by a factor of five, to $\sim 1.2\arcsec$ (0.3\,pixel) on
the celestial equator.

\ifthenelse{\boolean{emulateapj}}{
    \begin{deluxetable}{lr}
}{
    \begin{deluxetable}{lr}
}
\tablewidth{0pc}
\tablecaption{
	System parameters for the HATSouth project.
	\label{tab:specs}
}
\tablehead{
	\colhead{Parameter} & 
	\colhead{Value}
}
\startdata
\sidehead{\textbf{Telescope Mount and Dome}}
Initial positioning accuracy (DEC)				&	$\sim55\arcsec$\\
Initial positioning accuracy (RA)				&	$\sim140\arcsec$\\
Periodic error (peak-to-peak, no TDM)			&	$\lesssim 10\arcsec$\\
Periodic error (with TDM)\tablenotemark{a}		&	$0.5\arcsec$\\
Tracking error in 4 minutes 
(with TDM)\tablenotemark{b}						&   $1.2\arcsec$\\
Coarse motion speed								&	2.2\arcdeg/s\\
Stepper motor resolution						&	0.5\arcsec/step\\
Telescope home time (typical)					&	60\,s\\
Telescope home time (max)						&	200\,s\\
Dome opening/closing time						&	80\,s\\
\sidehead{\textbf{Optical Tube Assemblies}}
Clear aperture of primary mirror				&	180\,mm\\
Secondary mirror (projected diameter)			&	80\,mm\\
Focal ratio										&	2.8\\
Focal length									&	500\,mm\\
Focusing accuracy								&	$2\,\mum$\\
Filter											&	Sloan $r$\\
\sidehead{\textbf{CCDs}}
Chip											&	Kodak KAF 16803\\
Number of pixels								&	\ccdsize{4K}\\
Pixel size										&	$9\,\mum$\\
Full-well capacity								&	100,000\el\\
Gain											&	1.4\eladu\\
Readout noise									&	7.7\el\\
Cooling	with respect to ambient					&	$\Delta T = 45\,\C$\\
Dark current at $-30\C$							&	0.009\el/s\\
Readout time									&	25\,s\\
\sidehead{\textbf{Combined Instrument Parameters}}
Pixel scale										&	3.7\,\pxs\\
Field of view for single OTA					&	$4.18\times4.18\arcdeg$\\
Mosaic field of view 							&	$8.2\times8.2\arcdeg$\\
Vignetting (edge/corner)						&	67\%/46\%\\
Zeropoint magnitude (1\,ADU/s)					&	$r\approx18.9$\\
5-$\sigma$ detection threshold\tablenotemark{c}	&	$r\approx18.5$\\
Photometric precision at $r=11$\,mag			&	0.006\,mag/240\,s\\
Photometric precision at $r=13.3$\,mag			&	0.01\,mag/240\,s\\
Duty cycle\tablenotemark{d}						&	73\%\\
\sidehead{\textbf{Data flow}}
Raw compressed pixel data						&	19\,TB/year\\
Calibrated pixel data and photometry			&	$\sim60$\,TB/year\\
\sidehead{\textbf{Sites}}
Las Campanas Observatory, longitude				&	$70\arcdeg42\arcmin03.06\arcsec$ W\\
	~~~~Latitude								&	$29\arcdeg00\arcmin38.65\arcsec$ S\\
	~~~~Elevation								&	2285\,m\\
	~~~~Useful dark time\tablenotemark{e}		&	8.48\,hr/night\\
HESS site, longitude							&	$16\arcdeg30\arcmin10.17\arcsec$ E\\
	~~~~Latitude								&	$23\arcdeg16\arcmin23.32\arcsec$ S\\
	~~~~Elevation								&	1800\,m\\
	~~~~Useful dark time\tablenotemark{e}		&	7.15\,hr/night\\
Siding Spring Observatory, longitude			&	$149\arcdeg03\arcmin43.39\arcsec$ E\\
	~~~~Latitude								&	$31\arcdeg16\arcmin20.47\arcsec$ S\\
	~~~~Elevation								&	1165\,m\\
	~~~~Useful dark time\tablenotemark{e}		&	4.64\,hr/night\\
\enddata
\tablenotetext{a}{
	When there is no linear drift in the tracking speed e.g.~due to
	refraction. Periodic error is the displacement between the nominal 
	and actual position in RA while the mount is tracking.
}
\tablenotetext{b}{
	Including positional drifts.
}
\tablenotetext{c}{
	At a typical (median) sky background of $512$\,ADUs at LCO, 
	in a 240\,s exposure, for a source covering 12 pixels. 
}
\tablenotetext{d}{
	Fraction of time during a clear night with open shutter.
}
\tablenotetext{e}{
	Based on two years of weather data between 2010 March 15 and 2012
	March 15.  ``Useful'' means weather conditions met our criteria for
	opening as defined in \refsec{stat}. ``Dark'' means the Sun was below
	$-11\arcdeg$ elevation. 
}
\tablecomments{
	For more explanation on the data in this table see the main text.
}
\ifthenelse{\boolean{emulateapj}}{
    \end{deluxetable}
}{
    \end{deluxetable}
}

\subsubsection{The optical tube assemblies} 
\label{sec:ota}

We use Takahashi $\epsilon$-$180$ ED astrographs as our optical tube
assemblies (OTAs), which have a collecting area well matched to our
project goal.  They provide the fast focal ratio and high quality
optics we require at an affordable price.  Each HATSouth mount holds
four such Takahashi astrographs.  The $\epsilon$-$180$ astrograph is a
catadioptric system with an f/3.66 hyperbolic primary mirror of 180\,mm
clear aperture, a flat diagonal secondary mirror of 80\,mm diameter (as
projected along the optical axis), and an extra low dispersion
two-element Ross-corrector that flattens the field, reduces the focal
ratio to f/2.8, and reduces coma and chromatism.  The resulting focal
length is 500\,mm.  The aluminum optical tube is compact, with a length
of 568\,mm, an outer diameter of 265\,mm, and a total weight of
10.7\,kg.  The HATSouth mount was built to accommodate the combined
weight of the four OTAs and CCDs ($\sim70$\,kg) on the DEC axis.

\begin{figure}[!ht]
\plotone{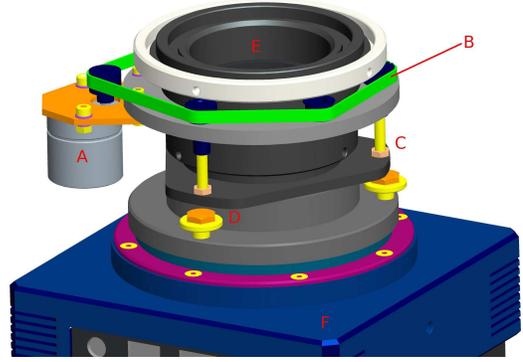}
\caption{
	Engineering model of the \hsfour\ focuser unit, replacing the
	original Takahashi focuser.  Displacement is driven through an
	tooth timing belt mating with small cogwheels on three
	fine-threaded screws that ensure symmetric driving.  Labeled are
	``A'': stepper motor, ``B'': sprocket, ``C'': fine-threaded drive,
	``D'': CCD tilt screws, ``E'': Ross-corrector, ``F'': CCD.
\label{fig:foc}}
\end{figure}

We removed the original Takahashi focusers from the OTAs, and replaced
them with custom-built focuser units (\reffig{foc}) that allow for very
fine electronic focus control via a two-phase stepper motor.  For
optimal image quality, the focal plane has to be fixed 77\,mm away from
the last vertex of the Ross-corrector lens (that is 56\,mm from the
Ross-corrector housing).  This means that the corrector-lens, the
filter, and the CCD form a single moving unit, which is moved
perpendicular to the optical tube when focusing.  One step of the
focuser corresponds to a motion of $2\,\mum$, and we make on the order
of 1000 steps (2\,mm) with the focusers during a typical night to
compensate for temperature changes, tube flexure and other effects. 
The moving unit is tightly held by three bearings, minimizing off-axis
wobble.  The force from the two-phase stepper motor is transmitted
through a tooth timing belt going around the entire focuser, and mating
with small cogwheels on three fine-threaded screws that ensure
symmetric driving (at 120\arcdeg\ offset) of the moving unit.  The
entire focuser unit is enclosed in a custom-made velvet sleeve to
prevent dust and other material getting onto drive mechanism or into
the focus unit itself.

The light leaving the Ross corrector first goes through a 5\,mm thick
$50\times50$\,mm Sloan $r$ filter ($\sim550$--700\,nm) manufactured by
Asahi Spectra.  It then passes through the two camera windows of the
CCD.  Each window is made of 1.5\,mm thick UV fused silica with a
broad-band (400--700\,nm) antireflectant (BBAR) coating.

As before, one prime consideration was to build a system that is
virtually maintenance-free, or at least, where maintenance is made
easy.  The Newtonian design is prone to collecting dust and unwanted
objects inside the tube, and is very laboursome to clean.  For example,
black mamba snakes ({\em Dendroaspis polylepis}), which are common in
Namibia, are notoriously hard to remove from a telescope tube.  Thus,
we sealed the front of the OTA by installing a flat optical glass.  We
used Schott B270 glass with 5.5\,mm thickness, 242\,mm clear aperture,
$<\!0.015$\,mm wedge, and transmitted wavefront error smaller than one
wavelength at 633\,nm.  The glass was coated for anti-reflection on
both sides with $R\!<\!0.5\%$ at 570--710\,nm.  The optical glass was
placed in a custom-built circular carrier, and can be removed easily
for cleaning by attaching a fitted handle to the outer rim and then
releasing three screws.

We also designed a dewcap that mounts on the front of the OTA,
primarily to decrease scattered moonlight.  The dewcap also contains a
low power (4W) heating coil which prevents dew condensing on the flat
optical glass.  The small amount of heat generated does not degrade our
image quality.

The large format of the CCD chips ($36.8\times36.8$\,mm) and the fast
focal ratio (f/2.8) necessitate accurate alignment of the chip normal
vector with the optical axis.  Even with perfect alignment, stellar
profiles on the edges and in the corners are slightly asymmetric, but
the variation of these profile parameters exhibits a symmetric pattern
with respect to the center of the field (e.g.~stars are elongated
perpendicular to the radial direction in all four corners of the chip). 
Without adequate alignment of the CCD normal and the optical axis, the
stellar profile parameters are asymmetric with respect to the center,
and in general are less circular, which adversely affects the focusing
stability and photometry.  Our focuser has three fine-alignment screws
that allow for manual alignment of the CCD (marked as ``D'' on
\reffig{foc}).  This is an iterative procedure, whereby through a
series of exposures we adjust the tilt of the CCD relative to the
focuser until the stellar PSFs in the corners of the CCD chip appear
symmetric.  This CCD alignment need only be performed once after
mounting the CCD to the focuser unit.  Our pixel scale is $\sim
3.7\pxs$, and stars that are in perfect focus have a Gaussian profile
with full-width at half maximum (FWHM) of $\sim2$\,pixels (7.4\arcsec). 
Our typical FWHM, averaged over the entire frame, is $\sim2.5$\,pixels
(9.2\arcsec).  We found that the vignetting (fraction of light detected
with respect to the center) in our system is $88\pm1$\% half-way to the
edge, $67\pm1$\% at the edge of the CCD, and $46\pm1$\% at the corner
of the CCD.

\subsubsection{The CCD Cameras}
\label{sec:ccd}

Each optical tube hosts an Apogee U16M CCD camera, which was selected
to give us a large format CCD with small pixels at an affordable price. 
While back-illuminated devices are known to be superior in performance,
acquiring 24 of them was completely outside our budget.  The cameras
are in a standard ``D09F'' housing with a custom chamber design that
has a slightly wider front opening to ensure that no light is blocked
at the corners of the CCD chip from the f/2.8 beam.  The camera has
three-stage Peltier (thermoelectric) cooling with forced air.  We
typically reach 45\C\ below the ambient temperature after $\sim30$\,min
cooling time and $\sim30$\,min stabilization time.

The CCD chip is a Kodak KAF16803 front-illuminated model. The
9\,$\mum$\ pixels have an estimated full-well capacity of $100,000$\el. 
We thus chose a gain setting of 1.4\,\eladu, so as to match the digital
and true saturation levels to $\sim60,000$\,ADU, which is just below
the $65,536$\,ADU range allowed by the 16 bit digitization.  The
average read-out noise for our 24 CCDs is 7.7\el.  The sensitivity
noise, which measures the relative sensitivities of different pixels
due to inhomogeneities of the chip, is $\sim0.02$.  These are mostly
corrected by careful flatfielding.  We measure a typical dark current
of 0.009\el/s at $-30$\C\ chip temperature (as derived from the median
of the dark pixels).

The chip features anti-blooming technology, preventing saturated pixels
from bleeding into near-by pixels.  While this has the advantage of
minimizing the area lost on the CCD due to over-exposed bright stars,
it also has the disadvantage of decreasing the quantum efficiency and
yielding less homogeneous sub-pixel structure (due to the anti-blooming
gates).  The pixels are illuminated from the front side, i.e.~from
towards the electrodes.  The pixels have a double structure, with one
half being polysilicon, the other half a transparent indium-tin-oxide
(ITO) layer.  There is a microlens above each pixel, directing light
preferentially toward the ITO gate, thus increasing the overall quantum
efficiency.

The CCDs are controlled via the USB-2.0 protocol, which has a maximal
cable range of 5\,meters.  To overcome this limitation we use an Icron
USB extender.  This extends the USB port of the computer through an
optical fiber to a remote hub with 4 USB ports, one for each of the
CCDs on the mount.  This solution has the additional advantage of
providing overvoltage protection through optical isolation (i.e.~using
light for coupling between the electronic components).

\subsubsection{The Dome}
\label{sec:dome}

The domes were designed and built by our team specifically for the
HATSouth project.  The design (\reffig{eng}) was based on the
asymmetric clamshell dome of HATNet.  This allows opening and closing
the dome in {\em any} position of the telescope, which is an important
consideration for robust automated operations.  For example, the dome
can close if the telescope mount breaks and is stuck at an arbitrary
position.  At the same time, the asymmetric clamshell design provides a
very compact dome size.
The dome hood is operated by a DC motor through a series of gears. 
Counterweights are used for balancing the dome hood, so little force is
required for moving the hood unless there is significant wind-load. 
The drive and structure are strong enough that we could safely operate
in windspeeds up to about 20\,\ms, but we close at 13\,\ms\ due to the
windshake on the OTAs degrading the stellar profiles.  When the dome is
fully open, the entire sky is visible down to an elevation of $\sim 10
\arcdeg$, except for towards the celestial pole, where this limit is
$\sim25\arcdeg$.  The telescope is hosted on a concrete pier that is
isolated from the dome so that windshake of the dome is not transferred
to the mount.  There are two water-proof limit switches for each of the
close and open positions of the dome.  Motion of the dome motor in a
given direction is inhibited if any of the relevant limit switches is
activated.

Over two years of operations the domes proved to be weather proof, with
no precipitation reaching the inside components, even under conditions
of torrential rain and high winds.  The sealing around the rim of the
dome is good enough to significantly decrease the concentration of dust
and the relative humidity inside when the hood is closed.  Also,
together with regular movement of the telescope, the dome has been
efficient in keeping out wildlife, such as insects, which is a serious
issue for the Namibian and Australian operations.

All domes have an 80\,W heater cable along the rim, which can be turned
on to eliminate the formation of ice at this critical surface; such ice
could prohibit closing the dome hood.  As an additional safety
consideration, the maximal current the dome motor is allowed to draw is
limited to 1\,ampere, preventing the dome mechanism from breaking, in
case the dome is stuck without reaching the relevant end positions. 
The domes are fitted with fans on the bottom panels that circulate air
through the interior in order to keep the temperature inside them equal
to the outside air temperature.

Each dome is fitted with a number of fail-safe mechanisms.  A Vaisala
DRD11A rain detector is attached to a console on the dome.  In case of
precipitation, the dome hood is forced to close, even if conflicting
commands are issued by the control computer.  Similarly, a photosensor
in a diffuse white sphere is attached to the same console.  If the
ambient light level is higher than that of the sky at sunset, the
photosensor commands the dome hood to close.  These fail-safe functions
can be disabled with an override switch in the dome electronic box. 
Finally, if the external power to the electronics is lost, it forces
the dome to close by drawing power from a 24\,V backup battery.  This
battery is constantly recharged when the dome is under power.

\subsubsection{Electronics}
\label{sec:elec}

\begin{figure*}[t]
\begin{center}
\plotone{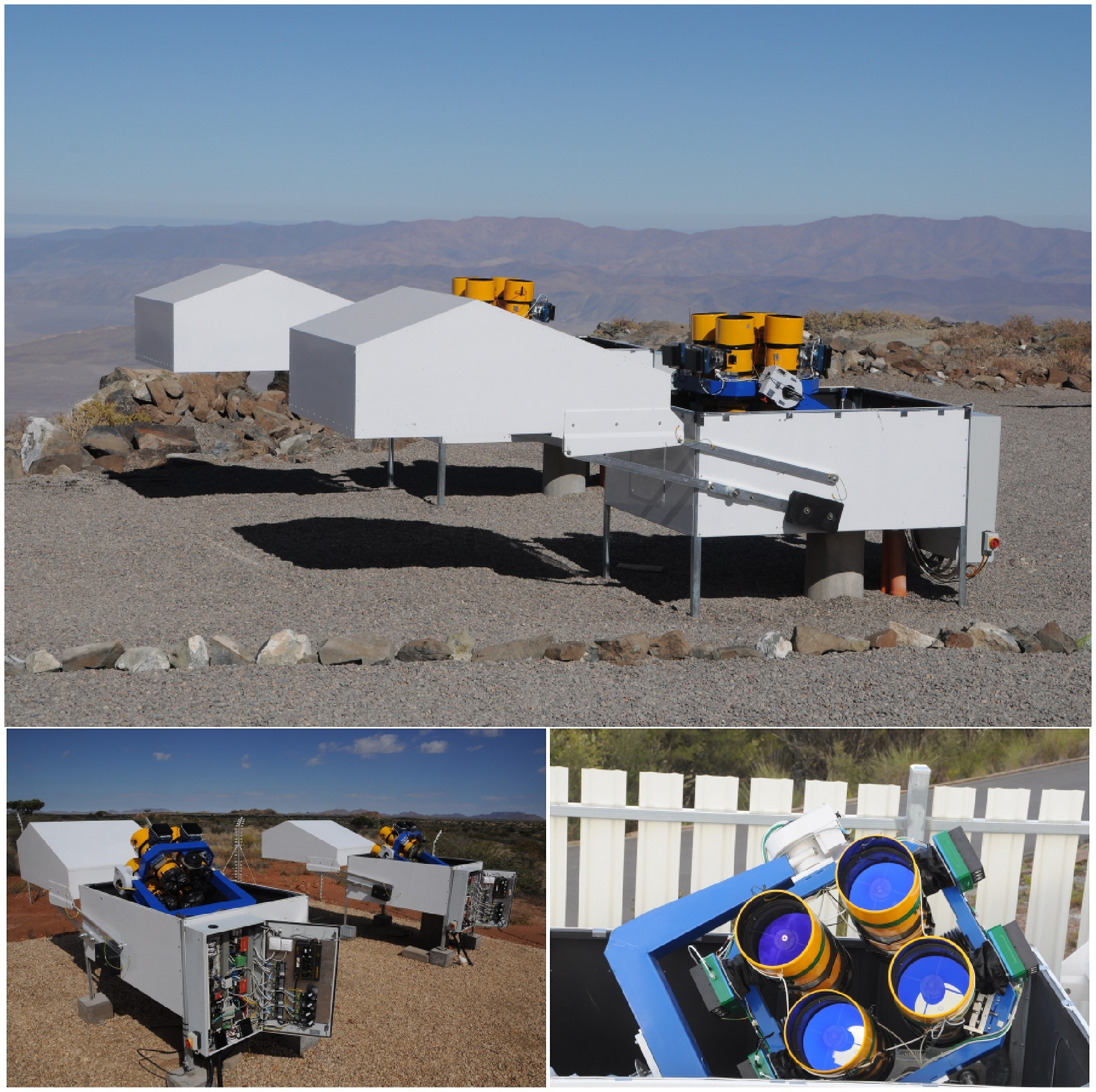}
\caption{
	The \hsfour\ units at Las Campanas Observatory, Chile (top), with
	the telescopes pointing towards the zenith.  The \hsfour\ units at
	the HESS site in Namibia (bottom left), with telescopes stowed
	towards the south celestial pole.  The electronic boxes are open
	showing the dome and telescope electronics described in
	\refsec{elec}.  A close-up view of an \hsfour\ unit at Siding
	Spring Observatory in Australia (bottom right), with part of the
	wind fence visible in the background.  \label{fig:stations}
}
\end{center}
\end{figure*}

Electronic components are housed in a weather proof,
$\mathrm\sim1\,m\times0.5\,m\times0.25\,m$ steel box that is attached
to the northern wall of the dome (these are visible in the lower left
panel of \reffig{stations}).  Cables originating from the near-by
control building reach the electronic box through a cable pipe; these
include printer port cables for the dome and telescope control, an
optical fiber cable for the CCD control and data download, and a
separate fiber for TCP/IP communication with other components (see
below).  A separate cable pipe leads two power cables to the electronic
box; one for powering the four CCDs and another for all other power
supplies.  The main (safety) switch on the electronic box cuts all dome
power, including that coming from the dome batteries.  However it does
not cut power to the CCDs so as to avoid a sudden warming of the
cameras.

At the heart of the electronic box is a modular programmable logic
controller (PLC) unit that is responsible for receiving signals from
the control computer and the various sensors on the dome and telescope,
and for issuing control signals to a wide variety of actuators.  The
PLC is a very robust, simple and compact industrial computer with a
high tolerance for extremes in temperature, dust, high humidity, etc. 
It is a hard real-time system, producing output within a very short and
well-known time interval after receiving and parsing input signals. 
The PLC is a common solution for industrial applications, especially in
cases where modifications to the system are expected (as opposed to
printed circuit boards with micro-controllers that are typically used
in mass-produced applications).

We illustrate the operation of the PLC with a few selected examples. 
If any of the dome open limit switches is activated, the PLC receives
this information, and using the embedded software, it interrupts the
motion of the dome motor, and inhibits any further motion in the open
direction.  If the control computer requests turning on the dome rim
heater cable, then the PLC turns on the relevant relay.  If the
telescope passes through the home proximity sensor, the PLC generates
an interrupt (IRQ) and sends it to the control computer via the
``scope'' printer port cable.

The software running on the PLC has been developed by our team.  It can
be uploaded (modified) over its network connection from a remote
location.  Of course, such a remote software upgrade is performed
through appropriate safety mechanisms.  Regarding the operation of the
telescope, the PLC receives the telescope RA, DEC and focus motor
instructions (direction and clocking signals) from the printer port
cable, and relays these commands to terminal stage cards that directly
control the motors.  Although the printer port is hardly used nowadays,
it is a good choice for low-level bi-directional communication, and for
generation of electronic control signals in the kHz range (such as for
driving the stepper motor) directly by the control computer.  We stress
that we have not implemented a full motion control in the PLC, such as
high speed coarse motion, ramping up, traveling a fixed distance,
calculating celestial positions, etc.  Instead, these signals are
calculated and transmitted by the control computer via the scope
printer port (see \refsec{swlow}).  The TDM units (\refsec{mount}) for
the RA and DEC axes also reside in the electronic box.

The electronic box has three separate 24\,V industrial power supplies:
one for the dome, another for the telescope mount, and the third one
for the PLC unit.  In addition, each of the four CCDs has its own 12\,V
power supply.  The power for the CCDs is fed through a network power
switch, which enables us to control their power remotely via TCP/IP.\@
Several other devices are attached to the network power switch, such as
a 4-channel thermometer measuring the telescope tube temperatures, the
Icron USB extender, and the electronic box thermometer.  The electronic
box is cooled by two strong fans that circulate air through filters,
which is critical for operations during the summer months.

The electronic boxes have two LED status lights mounted on the outside
of the door panel.  One LED indicates that there is power running to
the dome.  The other indicates that the \hsfour\ unit is operating, by
which we mean the that {\tt virtual observer} (see later) is in a
``run'' or ``weather-sleep'' state (see \refsec{swlow}).  These LEDs
are informative for any person at the site.  The status indicators are
also clearly visible in most conditions from the low-light web camera
mounted in the HATSouth control building (\refsec{wth}).  While the
status of these LEDs is directly accessible (and changeable) through
the control computer, it may happen that the control computer is
unreachable, and the web camera is used to assess the status of the
system.

\subsection{Weather sensing devices}
\label{sec:wth}

Reliable sensing of the current weather conditions is essential for
robust automatic operations.  At each site an array of weather sensing
devices are attached to the rooftop of the control building.  Data from
these devices are read by the node computer (see \refsec{compsys}).

A Vaisala WXT520 weather head measures wind speed and direction,
ambient temperature, precipitation, relative humidity and atmospheric
pressure.  The device has no moving parts as the wind speed and
direction are measured by ultrasound.  This is our primary source of
information on the wind speed, precipitation, and relative humidity. 
The device is connected via a RS-232 (serial) port on the node
computer, and is read through a text
based protocol.

A Boltwood Cloud Sensor II is used to establish the amount of cloud
cover.  This device compares the amount of radiation coming from the
sky (in a 150\arcdeg\ angle) with that coming from the ground, in the
8--14\,$\mum$ band \citep[for more details, see][]{marchant:2008}.  A
large temperature difference corresponds to cold (clear) skies, whereas
a low temperature difference corresponds to warm (cloudy) skies.  The
device is only moderately sensitive to high altitude cirrus clouds.  In
addition to providing a reliable measure of the cloud cover, it also
measures precipitation, wind speed, humidity, and the ambient light
level.  We read data from this device through the USB port of the node
computer.  We made use of the software library provided by {\tt
MyTelescope.com} in the data acquisition.  The connection to the cloud
sensor has surge protection, but no optical isolation.

Thunderstorms can form and move very quickly, especially at our
Namibian and Australian sites, and often lead to anomalies in the power
grid, increasing the risk of an instrument failure.  Sudden and intense
precipitation or lightning can also damage the instruments.  We use a
Boltek LD-250 lightning detector to monitor lightning storms.  This
detector is capable of measuring the direction and strength of the
strikes.  Large storm systems are easy to track when the lightning
strikes are displayed in polar coordinates (assuming observed strength
correlates with the inverse square of the distance).  The HATSouth
telescopes shut down if the storm level reaches a prescribed limit of
lightning strikes per minute.

Visible monitoring is always greatly reassuring to humans supervising
the HATSouth operations.  It can also reveal environmental conditions
that may otherwise go unnoticed.  Examples include objects left near
the clamshell domes that might impinge on them opening or closing,
nearby bush-fires causing a high density of smoke, and light pollution
from lights in surrounding buildings.  It can also help confirm the
veracity of the readings from the other weather sensing devices.  We
use an AXIS 221 Network Camera (version 4.45.1) mounted inside the
HATSouth control building and pointed though a glass window towards the
two \hsfour\ units to visibly monitor operations.  This camera works
well at low light levels, so can be used to monitor most night-time
operations.

An all-sky fisheye camera is installed at the Las Campanas and Siding
Springs sites.  This system, called CASKETT, is still under
development.  CASKETT uses a DMK 41AU02.AS CCD with a monochrome Sony
ICX205AL progressive scan chip that has no infrared cut filter.  It
produces a 180\arcdeg\ field imaged onto the $1280\times960$ pixel
CCD.\@ The exposure time is automatically adjusted based on the light
level, and the camera works day and night, and is not harmed by direct
sunlight.  We are currently working on software that reports the cloud
cover based on the CASKETT images, paying particular attention to high
altitude, cold cirrus clouds that are not robustly detected by the
Boltwood cloud sensor.

\subsection{Computer system}
\label{sec:compsys}

The HATSouth control building at each of the sites hosts the computer
system that is responsible for operating the HATSouth instruments.  We
have four computers at each site; one control computer for each of the
two \hsfour\ units, one node-computer, and a server.  All these are
mounted in a standard computer rack.

Each control computer manages an entire \hsfour\ unit, including the
dome, telescope mount, attached devices, and all four CCDs.  In
addition, the control computer performs real-time analysis of the
images acquired with its \hsfour\ unit, such as on-the-fly
calibrations, astrometry, point-spread-function (PSF) analysis,
focusing, and other tasks.  The control computers are rack-mountable,
and have a semi-industrial chassis with 4\,GB of memory, an AMD Phenom
9750 Quad-Core 2.4\,GHz Processor, and a RAID-1 array of operating
system and data hard-drives.  Communication to the dome and telescope
mount is via printer port cables that connect to a dual printer port
card through printer port overvoltage protectors.  The CCDs are
accessed via USB.  A serial port card is installed for connecting to
the uninterruptible power supply (UPS) units -- one for the computer,
and another for the \hsfour\ unit (dome, telescope, CCDs).  A watchdog
card executes a hard reset of the computer if the operating system
becomes unresponsive.  The control computers have been running
essentially non-stop for over two years.  Thanks to the RAID setup,
occasional failures of hard-drives did not affect operations, and the
faulty drives were swapped for new drives with no downtime or loss of
data.

The control computers run Linux Debian 6 operating system.  The kernel
has been patched with a real-time framework for Linux called Xenomai. 
This patch modifies the kernel to make it capable of executing certain
tasks in real-time, while taking care of other tasks at lower priority. 
For example, by using a special kernel driver that exploits the
advantages of Xenomai, we can issue clock signals (a periodic step
function) on the printer port of the computer at 16\,KHz, and drive the
two-phase stepper motor of the telescope mount without experiencing
glitches due to sudden changes in the frequency or the width of the
clock signals.  Such glitches would not only lead to imprecise
pointing, but would also damage the telescope mount as the motor loses
sync during the motion.  Another example is that we can track the
telescope mount with 30.08\,Hz to achieve sidereal rate tracking (since
1 microstep of the motor is 0.5\arcsec).  In the meantime, the computer
is still running in a multi-task mode, and is managing a huge number of
other processes such as ethernet communication, data processing, CCD
control, authentication of log-ins, network firewall, etc.  We have
developed Xenomai kernel drivers for controlling the telescope and the
dome.

The node computer has basically the same hardware and architecture as
the control computer, and is responsible for a number of important
functions.  It monitors the weather sensors (\refsec{wth}) and stores
the information in a MySQL database (see \refsec{swlow}).  It also
receives accurate time from our Garmin 16x HVS GPS unit that is mounted
to the control building roof.  Data from this GPS device is read
through a serial port of the node computer via the {\tt gpsd} daemon. 
We use the pulse per second (PPS) sharp time synchronization signal,
which improves the time accuracy up to the IRQ response time of the
serial port.  The GPS shows up on our node computer as a ``stratum-0''
network time protocol (NTP) reference, and measures time in UTC.  The
node computer also monitors the status of the UPS units that power the
weather sensors and internal electronic devices (web camera, signal
converters, network switches).  Finally the node computer hosts the
webpage containing all the weather information for that site (see
\refsec{stat}).  Since no real-time tasks (instrument control) run on
the node computer, the Linux kernel is not patched with Xenomai.

A server computer is used for storing data on the site, performing data
processing that is not related to real-time reductions, transferring
data via the Internet, and archiving data to tapes.  We use an
Ultrium-4 tape drive to archive tapes with 800\,GB capacity.  The
server has 2 Quad-Core AMD 1.9\,GHz Opteron processors, 32\,GB of RAM,
and 10\,TB of local storage on a RAID-6 array of disks.  Such a buffer
on the site guarantees that operations are never halted due to lack of
disk space and potential delays in shipping of the tapes.

The computers are connected via a dedicated internal network, and are
connected to the Internet via another network switch.  A number of
additional devices are part of the computer system, such as an internal
web camera allowing remote monitoring of the control building, an
external low-light web camera that is pointed at the telescopes
(\refsec{wth}), signal converter and transient isolating units for the
weather devices and the GPS, and remotely manageable power switches to
power cycle equipment.

All computers and electronic devices are connected to UPS systems.  In
the case of a short power failure ($\!<\!30$\,s) the UPS systems allow
operations to continue without interruption.  For longer power
failures, the system is cleanly shut down, including closing of the
\hsfour\ units, halting the computers, and at the very end of the
procedure, turning off the UPS units to avoid complete discharging.  If
the power returns, the UPS units will wait until they are sufficiently
charged, and then the systems start up automatically.

The HATSouth Data Center (HSDC) is located in the server room of the
Princeton Institute for Computational Science and Engineering (PICSciE)
at Princeton University.  The HSDC consist of a number of server
computers running Linux that our group manages.  These server computers
typically have 32 CPU cores, 20--40\,TB storage space, and 64\,GB
memory.  It is here that all the data from the three observing sites
are collated and the bulk of the processing occurs.  The data flow and
reduction that occurs at the HSDC is set out in \refsec{dr}.

\section{The HATSouth instrument control software}
\label{sec:csw}

As described in \refsec{compsys}, each \hsfour\ unit is controlled by a
single control computer running Linux with a special kernel that is
capable of real-time operations.  In addition, a node-computer is
responsible for weather sensing and synchronizing the time to the GPS
time.  A large suite of software is running on the control and node
computers, responsible for the instrument control.  We broadly classify
the control software components to ``low-level'', meaning direct
control of instruments, and ``high-level'', referring to more general
observatory control, usually connected to the ``low-level'' software.

\subsection{Low level software}
\label{sec:swlow}

\subsubsection{Scope}

The control of the telescope mount is performed through a Xenomai-based
(real-time) character device driver, called the \scope\ module.  This
module depends on the basic built-in printer port control modules of
Linux ({\tt parport, parport\_pc}).  When the \scope\ kernel module is
loaded, a number of initial parameters are supplied, such as the choice
of the hemisphere (to determine the direction of tracking), the
resolution of the axes (e.g.~0.5\arcsec/pix), settings for ramping up
the motors to maximal slewing speed, and the level of verbosity.

The telescope is represented by two files.  The first one ({\tt
/dev/scope}) is used for issuing commands to the mount, simply by
echoing the relevant commands into it.  For example, {\tt echo home >
/dev/scope} initiates the automatic homing procedure of the mount.  The
second file ({\tt /proc/scope}) is for reading the status of the mount,
showing the detailed status of the RA and DEC axes, the focus motors,
and the TDM.  This is a very robust solution, whereby the control is
done through a kernel driver, and is running at much higher priority
than the user-space programs (such as a browser).  In addition, the
{\tt ioctl} (input-output control) mechanism is used for certain
operations, such as aborting the current activity of the mount.

\subsubsection{Dome}

Control of the dome is similar to that of the telescope mount (via the
\scope\ driver, see above), and is performed through the {\tt dome}
kernel driver.  A separate printer port is used for controlling the
dome.  Basic operations include turning the power on/off, closing or
opening the dome hood, and controlling the dome heating, cooling fans,
and dewcap heaters.  The status of the dome is read through the {\tt
/proc/dome} file.  For example, the status of the dome hood can be
``open'', ``closed'', ``unknown'' (no limit switches activated), or
``error'' (at least one open and one close limit switch is activated at
the same time, indicating an electronic or mechanical failure).

The \dome\ and \scope\ kernel drivers have been running very robustly
for over 2 years, with not a single case of computer failure traced
back to kernel driver errors.  Older versions of these drivers have
been running on the HATNet project for 8 years.

\subsubsection{CCDs}

As described in \refsec{ccd}, the four cameras on a single mount are
all connected to the USB bus of the control computer via an Icron fiber
extender.  Since the CCDs have the same USB identifier, they are
instead identified by our software reading out their serial numbers. 
The camera control is based on the software library supplied by the
company RandomFactory (David Mills).  We made minor modifications to
these codes, and developed a camera server ({\tt ccdsrv}) on top of
them that is capable of the simultaneous control of multiple CCDs. 
Also, it is compatible with our higher level observatory control and
existing data structures, such as our required FITS header keywords and
loading configuration parameters from a MySQL database.

\subsubsection{Weather devices}
The status of the Vaisala weather-head and the Boltek lightning
detector are read through the serial ports of the node computer.  The
status of the cloud detector is read through the USB port.  Each device
has a separate, custom-developed, software daemon (code running in the
background in an infinite cycle) that is responsible for these
operations.  Weather information is read out from the detectors every
30\,seconds.  The daemons use a simple text based communication
protocol over TCP/IP (e.g.~addressable by {\tt telnet}).  All sensor
reads are automatically logged into a local MySQL database host on the
node computer.

\subsection{High level software}
\label{sec:high}

\subsubsection{Mount-server}

At the bottom of the higher level codes is the mount-server
(\mountsrv), which runs on the control computer, and allocates the
\dome\ and the \scope\ kernel drivers, so that commands to these
drivers are only accepted through the \mountsrv.  This safety mechanism
avoids competing commands issued to the hardware.  The \mountsrv\
separates incoming commands (e.g.~open the dome, move the telescope),
and channels them to the relevant device.  The \mountsrv\ communicates
with higher level codes via TCP/IP connections.

\subsubsection{Weather Daemon}

Another important daemon is {\tt wthdaemon}, which runs on the node
computer.  This listens to all the individual weather devices.  The
{\tt wthdaemon} establishes if the weather conditions are suitable for
observing based on the limits set out in \refsec{sites}.  In addition
to these weather limits, the {\tt wthdaemon} imposes time-outs of
20\,minutes for clouds, high windspeed or high rate of lightning
strikes, and 60~minutes for rain, hail and high humidity.
These time-outs ensure the domes do not open and close repeatedly in
marginal weather that is close to our limits, and provides for a
measure of conservatism appropriate for fully automated operations.  If
the {\tt wthdaemon} reports suitable conditions and no time-outs, and
if the telescopes are assigned weather dependent night-time tasks, they
will open up and execute those tasks.  The weather conditions are
logged in a MySQL database, and thus their previous values are known
even if the software or the computer is restarted.

\subsubsection{Virtual Observer}

The most significant high level software is our ``{\tt virtual
observer}'' (\vo), controlling all the hardware in an optimal manner
through the lower level software described above, and making
intelligent decisions based on the conditions.  The \vo\ is an
idealized observer, running in an infinite loop, always being alert of
the conditions, constantly trying to keep operations optimal, and
always having an oversight of the priorities.  The \vo\ is connected to
the \mountsrv, the \wthdaemon, the \ccdsrv\ and the MySQL database.

In addition to being an infinite loop, the \vo\ has four separate
internal states.  If there are no tasks defined that could be executed
(typically during the day-time), the \vo\ idles in ``daysleep'' state. 
The CCDs are warmed back to a temperature around 0\C, the dome hood is
closed, the telescope points to the celestial pole to avoid pointing at
the Sun should the dome be opened.  While in daysleep, the \vo\
periodically checks if anything changed, e.g.~an observing task has
been defined that requires preparation of various hardware devices. 
Note that observing tasks need not necessarily be carried out in the
night time with an open dome, e.g.~dome-flats can be taken at broad
daylight with the dome closed.

If such a task is found, the \vo\ changes into ``run'' state, and
prepares the devices, most typically cooling down the CCDs to operating
temperature (see \reftab{specs}), and moving the telescope to its home
position.  The task is then executed based on a priority system, and
the observer stays in ``run'' state as long as there are tasks to be
completed and the conditions for these tasks are appropriate.

If the task to be performed requires good weather (basically anything
that assumes an open dome hood), but the weather conditions are
adverse, the \vo\ transitions into the ``weather-sleep'' state.  Here
it waits for the weather conditions to improve, or the scheduled finish
time of the task is reached.  All devices are prepared for the
observations (CCDs are kept at low temperature for imaging) to enable
rapid transitioning to the ``run'' state, should conditions improve or
a task with no weather dependency appear.  If there are no current
tasks, the \vo\ transitions into ``daysleep'' state.

Finally, there are a number of semi-critical conditions under which the
\vo\ is forced to switch into ``suspend'' state by the ``big-brother''
software (see below).  In suspend state the observer idles, waiting for
the conditions to change back to normal.  Examples are: i) the station
loses connection to the outside world, ii) time is not synchronized to
UTC via the GPS or the Internet, or our system time is more than 0.1
second off the time standard, iii) connectivity to the UPS system is
lost, iv) there is no free disk-space, v) the health of disks or the
RAID arrays is critical.

In practice, the \vo\ is in the above infinite loop, in one of the four
states, for months at a time.  The \vo\ exits this loop in case of a
shutdown, or when we need to perform maintenance.  In the latter case,
automatic start-up of all hardware components is prohibited until this
``service'' state is cleared, so as to ensure safety of the personnel
performing the maintenance.

The \vo\ has a {\em very basic} capability of scheduling tasks.  We
have not addressed the complex problem of queue scheduling with
multiple institutions or principal investigators
\citep[e.g.][]{tsapras:2009}.  Tasks can have well defined start times,
weather dependency and priorities.  We have four distinct tasks,
descriptively named: \bias, \dark, \skyflat\ and \monf\ (which is
night-time science field monitoring).  These four tasks are launched by
the \vo, and governed through socket communication.  While a task is
running, the \vo\ is performing its standard activities.  For example,
in the case of bad weather, the \vo\ instructs the running field
monitoring program to cease operations, and then prepares the hardware
for adverse weather conditions (e.g.~it closes the dome hood).

\subsubsection{Big Brother}

``Big Brother'' (\hatbb) is a high level software component watching
the operating system and the rest of the HATSouth control software.  We
run \hatbb\ on the control computers and the node computer.  It
routinely checks the status of crucial operating system level services
({\tt ssh, ntp, mysql, gpsd}) and HATSouth control components ({\tt
observer, mountsrv, ccdsrv, wthsrv}, low level weather device daemons),
and if one is found to be not running or malfunctioning (e.g.~not
responsive), then the relevant service is restarted.  In case of low
disk space, {\tt hatbb} sends warning emails, and eventually turns the
system in suspend state.  Connection to the site computers is regularly
checked in the following way.  Automated services run on selected
computers at our Princeton-based HSDC, and connect to the site
computers via {\tt ssh} at regular intervals.  Should \hatbb\ running
at the site realize that no such connections have been made for over 3
hours, the HATSouth system is again pushed in service state. 
Similarly, if the status of the ethernet interfaces is not satisfactory
(interfaces are down, or show too many failed packets), the status of
the UPSes is not acceptable (no connection, batteries discharged), our
computer time is off by more than 0.1\,seconds, or the jitter on the
NTP time-stamps exceeds 0.1\,seconds, then \hatbb\ switches the {\tt
virtual observer} to suspend.  There are a number of serious error
conditions, for which, instead of ``suspend'', the system is sent in
``service'' state.  Such conditions are e.g.~if the dome is open in the
daytime, or during rain, or the dome driver showing an error status.

\subsubsection{The database}

The usage of configuration or log files has been minimized on HATSouth. 
Instead, configuration parameters and logs are kept in database tables. 
We use the free MySQL engine for this purpose. 
There is a central database on a selected server computer at our
Princeton HSDC with altogether 75 tables, 48 of which describe the
configuration of the network (``config''-type), 25 are for various
observation related logs (``log''-type), and 2 additional tables have
special functionality.  This database has local copies on the site
computers.  The telescopes always use the local (on-site) versions of
the central database, because the Internet connectivity between the
HSDC and the sites can be unpredictable, and operations should not be
slowed down by potentially slow connection with Princeton.

Any change in the configuration is first implemented in our central
database at the HSDC, and then synchronized to the individual databases
at the remote sites.  The telescopes are then instructed to recognize
the changes, and operate with the new configuration parameters.  For
example, if the pointing model is re-calibrated on a mount, a new
pointing model version is introduced in our central database, the table
is synchronized to all sites (with a single command), and the telescope
pointing model is changed on-the-fly.  Examples for configuration
tables are: i) identification and calibration parameters of the 24
CCDs, ii) map of the current setup, matching the station identification
(ID) numbers with the mount IDs, camera IDs, etc., iii) pointing models
for the 6 mounts, iv) setup of fields monitored by the \monf\ task,
etc. Similarly, all operation logs are kept in database tables on the site
computers, and are regularly synchronized back to the central database
at the Princeton HSDC.  The database structure and the scheme of
synchronizations is such that configurations or logs are never
overwritten.  This centralized setup is very convenient; the
configuration and logs of the entire network can be reviewed at one
location.  
For example, deriving statistics on how many frames were
taken by the HATSouth network on a given field is a matter of a simple
query in MySQL.

\section{The HATSouth sites and operations}
\label{sec:sitop}

\subsection{Observing Sites}
\label{sec:sites}

\begin{figure*}[t]
\begin{center}
\plotone{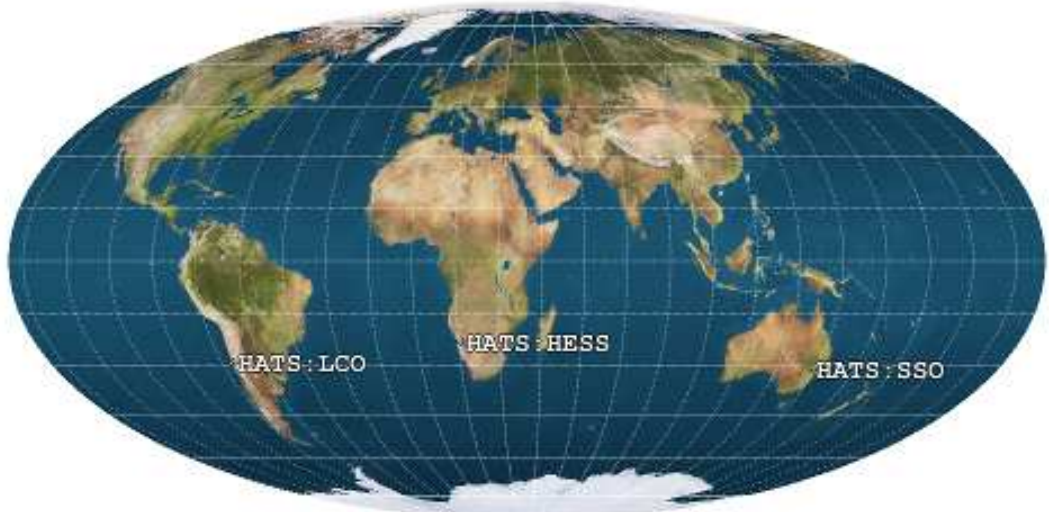}
\caption{
	Geographical location of the HATSouth sites on a Mollweide
	projection.  These site provide near-optimal longitudinal
	separation, with LCO $\rightarrow$ SSO = $141\arcdeg$, SSO
	$\rightarrow$ HESS = $133\arcdeg$, and HESS $\rightarrow$ LCO = $
	86\arcdeg$.  The three sites enable round-the-clock monitoring of
	selected southern stellar fields (image generated by {\tt
	xplanet}).
\label{fig:xplanet}
}
\end{center}
\end{figure*}
\renewcommand{\textfloatsep}{-4mm}

\renewcommand{\textfloatsep}{-4mm}

The HATSouth network is operating at three premier astronomical sites
in the southern hemisphere (\reffigl{xplanet}) that have a longitude
coverage allowing for round-the-clock observations of nearly any
celestial field in the southern hemisphere (that is close to the
anti-solar point).  The \hsfour\ units and control buildings were
installed at all three sites in 2009, and the operations during 2009
predominantly involved commissioning of the network.

For the past two years, we have been continuously monitoring the
meteorological conditions at the sites with 30\,second time resolution
using our weather sensing devices (\refsec{wth}).  Of primary interest
to us is the number of astronomically useful hours.  By this we mean
any time during the night when the weather server reports suitable
observing conditions {\em for our purposes}.  Based on monitoring
during our commissioning period, suitable observing conditions exist
only when:

\begin{enumerate}
\item The Sun elevation is below $-11\arcdeg$.
\item The sky temperature as measured by the Boltwood Cloud Sensor II
is below $\sim-30\C$, indicating either cloud-free conditions or only
thin, cold cirrus clouds.  (The exact value has been adjusted depending
on the site and the season.)
\item The average wind speed, as measured by the Vaisala weather-head
in 30\,s intervals, is below $13$\,\ms, in which case the amplitude of
windshake will have a negligible effect on our image quality.  Also, no
gusts exceeding 18\,\ms occur.
\item The relative humidity as measured by the Vaisala weather-head is
below 90\%, which safeguards against dew condensing on the telescope
front glass.  Once the humidity exceeds 90\%, it has to drop below 88\%
to declare the conditions suitable again.
\item The rate of lightning strikes as detected by the Boltek lightning
sensor is below 50\,strikes/min within 450\,km of the site and less
than 30/minute within 75\,km, indicating there are no major electrical
storms close to the site.
\item The precipitation rate as measured by the Vaisala weather-head is
zero, indicating no rain, hail or snow.
\end{enumerate}

We do not monitor atmospheric seeing, as our optical system delivers a
stellar PSF of $\sim 10\arcsec$, which is much wider than the typical
seeing at these sites ($\sim 1\arcsec$).  We note that our criteria for
suitable observing conditions relate critically to the specifics of our
hardware and project.  For example, other telescopes on the site may be
able to operate at windspeeds and humidity levels above the HATSouth
limits.  Below we describe the three HATSouth sites, including their
weather statistics (\reffig{obsstat}).

\paragraph{Chile} 
The Las Campanas Observatory (LCO) site
($70\arcdeg42\arcmin03.06\arcsec$ W, $29\arcdeg00\arcmin38.65\arcsec$
S) is located 110\,km north-east of La Serena, Chile, and is operated
by the Carnegie Institution for Science.  LCO is famous for its
extraordinary astronomical conditions, and hosts renowned telescopes,
such as the twin 6.5\,m Magellan telescopes, and the OGLE project, and
will be the site of the future 25\,m Giant Magellan Telescope (GMT). 
At 2285\,m elevation, it has a dark sky and good transparency.  The
speed of the current network link requires us to manually ship most
data back to the HSDC via Ultrium-4 800\,GB tapes.  There is very
little seasonal pattern in the fraction of useful (clear) hours; the
site is basically clear for most of the year.  The yearly average of
dark hours (Sun elevation $<-11\arcdeg$) is 10.2\,hours.  Based on two
years of weather data (from 2010 March 15 until 2012 March 15), there
are $\sim 8.48$ useful HATSouth observing hours per 24-hour period. 
Just 14\% of the dark hours are cloudy, and only a further 3\% of them
not useful for reasons other than cloud (primarily high humidity or
windspeed).  The first two \hsfour\ units were installed at LCO in 2009
May, and operations started in 2009 July.

\begin{figure}[t]
\begin{center}
\plotone{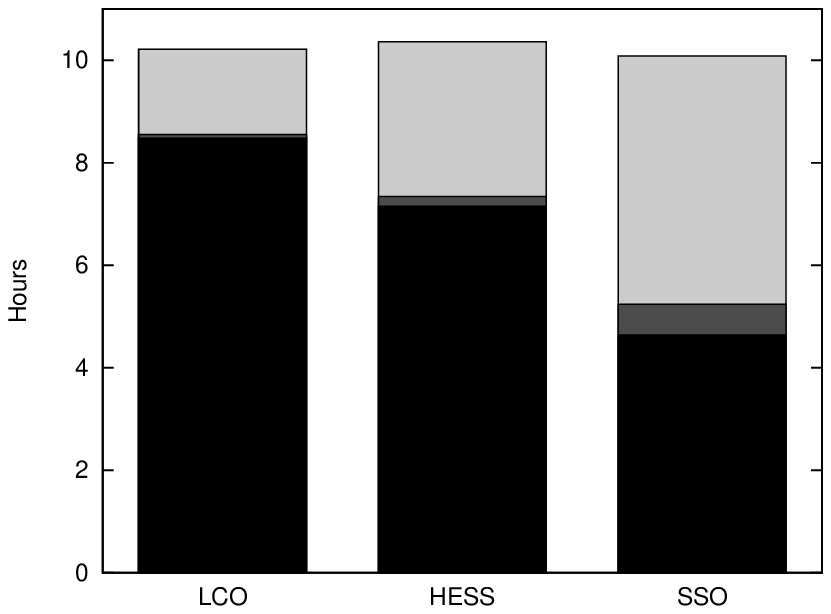}
\caption{
	Average number of dark hours for each site, broken down to useful
	(black), cloudy (light grey) or cloud-free but not useful (dark
	grey) due to other conditions (high humidity, wind, or lightning
	rate).  \label{fig:wthpie}
}
\end{center}
\end{figure}
\renewcommand{\textfloatsep}{-4mm}

\paragraph{Namibia}
The High Energy Spectroscopic Survey (HESS) site is located in the
Khomas Highland (1800\,m elevation) of Namibia, about 100\,km south
of Windhoek 
($16\arcdeg30\arcmin10.17\arcsec$ E,
$23\arcdeg16\arcmin23.32\arcsec$ S).
Max Planck Institute for Astronomy (MPIA, Heidelberg, Germany) has
access to the site through its partner institution, the Max Planck
Institute for Nuclear Physics.  Since the HESS project operates
atmospheric Cherenkov detectors, the site was chosen for its very dark
skies and good record of clear nights.  There is a wet season from
January through April, when the fraction of clear hours is
significantly less than during the rest of the year.  On average, we
have $\sim 7.15$ useful HATSouth observing hours per 24-hour period at
the HESS site.  26\% of the dark hours are cloudy, with a further 5\%
not useful for reasons other than cloud (primarily high humidity or
high lightning strike rates).  We have a dedicated satellite-dish based
Internet connection to the site with guaranteed 512\,Kbit\,$\mathrm
s^{-1}$ upload and download speed.  This is marginally sufficient for
monitoring the instruments and downloading the basic diagnostic plots. 
Similarly to the LCO site, data is shipped back via Ultrium tapes.  The
\hsfour\ units were installed to the site in 2009 August, and
operations started shortly thereafter.

\paragraph{Australia}
Siding Spring Observatory (SSO) is located in rural New South Wales in
Australia
($149\arcdeg03\arcmin43.39\arcsec$ E,
$31\arcdeg16\arcmin20.47\arcsec$ S).  
The site is owned and operated by the Australian National University's
(ANU) Research School of Astronomy and Astrophysics.  SSO is the
premier site for optical and infrared observation in Australia; it has
excellent dark skies and substantial infrastructure, including a fast
Internet link dedicated to research and training activities.  Several
Australian and international telescopes are located on the mountain,
including the 3.9\,m Anglo-Australian Telescope and the ANU's 2.3m and
SkyMapper telescopes.  The ANU has permanent on-site technicians who
can assist us if manual procedures are required.  There is no strong
seasonality in the fraction of clear hours.  The typical wind
conditions at the selected location for the HATSouth instruments
necessitated building a wind-fence around the telescopes.  This reduces
windshake on the telescopes and ensures the dome hood can open and
close without significant wind-load.  High humidity, with otherwise
clear skies and no wind, is more frequently an issue at SSO than at the
LCO and HESS sites.  The SSO site has had $\sim4.64$ useful hours per
24-hour period over the past two years.  49\% of the dark hours are
cloudy, with a further 5\% not being useful due to reasons other than
cloud (primarily due to high humidity over the winter months).  There
is not much annual variation in the distribution of clear nights.  We
note that the current two-year statistics is certainly biased: based on
annual rainfall records of the Australian Bureau of Meteorology, 2011
was the wettest year in Australia in the {\em last century}, and 2010
was among the top 10.  The Internet connection between the HSDC in
Princeton and SSO is very fast, reaching 40\,Mbit/s both ways, which is
high enough for us to transfer image data directly over the Internet
without the need to ship tapes.  The \hsfour\ units were installed in
2009 November, and operations started in 2009 December.

\subsection{Scheme of nightly operations}
\label{sec:oper}

The \hsfour\ units operate in a fully autonomous manner based on the
\vo\ ({\tt virtual observer}) software (\refsec{swlow}).  The \vo\
leaves its ``daysleep'' state an hour before sunset, and upon entering
the ``run'' state, it starts cooling the CCDs and homes the mount. 
Whenever the photosensor permits, the \vo\ opens the dome.  The
increased air-flow helps reaching a cooler set temperature for the
CCDs, which in turn reduces CCD dark noise.  The cameras are then kept
at this fixed temperature during the entire night regardless of changes
in the ambient air temperature so as to maintain CCD stability and
ensure appropriate calibration frames are available for data reduction.

The \vo\ starts the \monf\ task (our field monitoring program) when the
Sun is at $-11\arcdeg$ elevation.  All useful time between dusk and
dawn (Sun elevation $<-11\arcdeg$) is spent executing the \monf\ task. 
For the purpose of selecting stars to monitor with HATSouth, we
established a tiling of the entire sky consisting of 838
non-overlapping ``nominal'' fields, each $8\degr \times 8\degr$ wide. 
The HATSouth FOV is slightly larger than these fields, so by surveying
them we gradually map the entire sky.  We assign high priority to
$\sim12$ fields each year.  These ``primary'' fields are chosen for
observation based on several factors such as optimal visibility at the
given time of the year, the expected sky density of dwarf stars,
proximity of the field to Solar System objects, and prior history of
observations.  If none of the primary fields are visible for some
reason, we select a field from a list of ``secondary'' fields.  Our
visibility calculations include distances from the Sun, Moon and bright
Solar System planets, and constrain the zenith distance to be smaller
than 60\arcdeg.

Stability is of prime importance for maintaining high precision
photometry.  We make an effort to stabilize the positioning of the
sources (astrometry), and the PSF of the sources (via focusing). 
Certain effects, however, are unavoidable; for example the differential
refraction within the 8.2\arcdeg\ field is 32\arcsec\ at a zenith
distance of 60\arcdeg.  The stellar profiles may also vary due to
wind-shake.

\subsection{Astrometric Stability}
After each 240\,s science frame we run ``quick'' astrometry on all four
frames (originating from the four CCDs on the \hsfour\ unit), and
adjust the position of the mount to track on the nominal field center. 
We typically achieve this to within 10\arcsec (rms).  Astrometry is
performed using the Two Micron All Sky Survey
\citep[2MASS;][]{skrutskie:2006} catalog retrieved around the nominal
position and a list of stars extracted from the image.  The
transformation between the two catalogs is determined using triangle
matching, as described in \citet{bakos:2004} and \citet{pal:2006}. 
Should this astrometry fail, we fall back on using {\tt astrometry.net}
\citep{lang:2010}, which can produce a robust (but less accurate)
solution even for very large pointing errors.  Failing astrometry is
also an excellent diagnostic of various errors in automated operations,
ranging from broken motors to poor initial orientation to the dome not
opening.  One unexpected effect we have noticed is ``breathing'' of the
top frame holding the four telescope tubes, in the sense that the
relative pointing of the four OTAs is a function of the hour angle
(with an amplitude of $\sim20\arcsec$).

\subsection{PSF Stability}
\label{sec:psf}
The stability of stellar PSF is maintained using a series of
procedures.  The encoder feedback and the RA TDM (\refsec{mount})
ensure accurate tracking, with no more than $\sim1.2\arcsec$\
positional change over four minutes exposure.  Significant fork flexure
at high hour angle can result in slightly elongated stars (primarily in
the DEC direction), as the fork ``unflexes'' during the four minute
exposure.  We have quantified this effect for each \hsfour\ mount, and
compensate for it by stepping the mount in DEC during the exposures. 
Also, we have developed an autofocusing routine to compensate for focus
changes, due primarily to thermal effects adjusting the distance
between the CCD cameras and the OTAs.  First we take a series of frames
with short (30\,s) exposures, and adjust the focus in between the
exposures.  We then derive a map of the PSF parameters for each frame
in this focus-series, and establish which frame exhibits the best
focus.  This is a non-trivial task due to the field curvature, CCD
alignment, and the resulting variable PSF parameters across the field. 
We derive such a focus series once every two months.  During normal
field monitoring observations, we take a 30\,s focus-frame every 3rd
exposure.  We derive the PSF map for the focus frame, and find the best
match from our focus series.  We then adjust the focus counter with the
difference between the counter values for the best match and the
originally established best focus.  Short (30\,s) exposures are
required in order to minimize tracking errors, and fork flexure, and to
decrease the effects of possible wind-shake, both for the initial focus
series and for the regularly taken focus frames.  The focus frames are
not discarded, but instead are used for monitoring brighter stars that
exceed their saturation limit by $\lesssim 2.3$\,mag in the regular
240\,s science frames.

The \monf\ task runs all night, taking 4 minute exposures, followed by
$\sim 25$\,s readout time and $\sim 12$\,s time for astrometry and
autofocusing.  At the start of nautical twilight in the morning (Sun at
$-11\arcdeg$ elevation) the \vo\ starts taking skyflats in pre-selected
regions close to the zenith, but always at least 40\arcdeg\ away from
the Moon.  The \skyflat\ task is terminated at around sunrise, at which
point the dome is closed, and the \vo\ launches the \dark\ task which
takes a series of 20 dark exposures of 240\,s each.  Finally the \vo\
concludes the run by starting the \bias\ task, which takes a series of
20 bias frames.  The \vo\ then enters ``daysleep'' state, the CCDs are
warmed up to $\sim 0$\C (because in the daytime they would not be able
to maintain the set temperature), and the \vo\ waits for the the start
of the next night.

\begin{figure*}[!ht]
\plotone{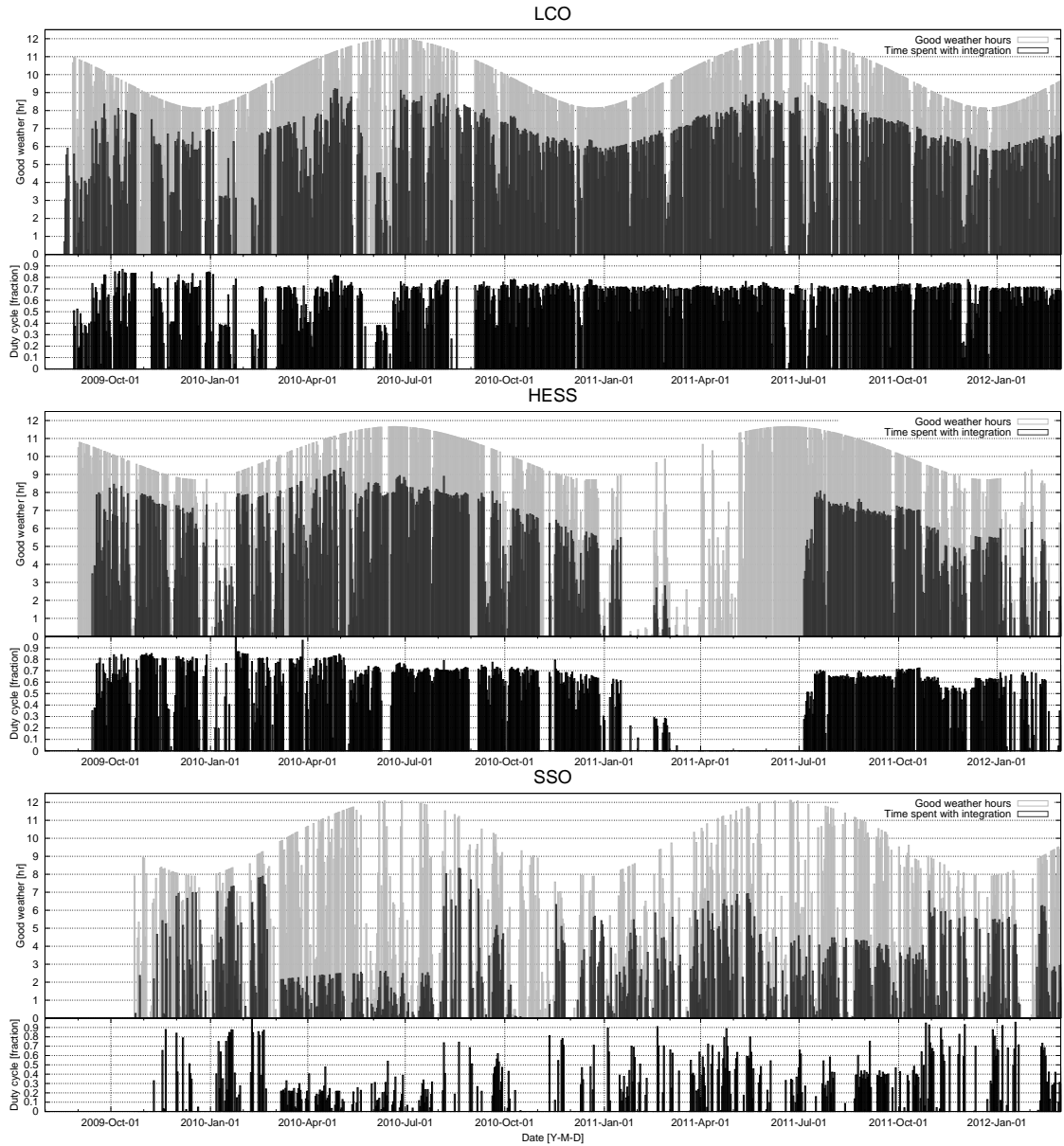}
\caption{
	Observing statistics from the LCO, HESS and SSO sites.  The top
	panel shows in light gray columns the good weather night-time hours
	for LCO as a function of date, since 2009 September.  Dark gray
	columns show the actual hours spent with open shutters, averaged
	over the 8 CCDs per site.  The panel immediately below shows the
	duty cycle, i.e.~the ratio of the dark gray to light gray columns. 
	The third and fourth panels show the same information for the HESS
	site, and the fifth and sixth panels refer to SSO.
\label{fig:obsstat}}
\end{figure*}

\subsection{Manual Interaction with the Telescopes}

A number of software tools are available for manual interaction with
the HATSouth control software components, notably with the {\tt virtual
observer}.  These assume opening a secure shell ({\tt ssh}) connection
to the control computer.  The software components can be started,
stopped, restarted, or checked for operation.  The weather status can
be declared as inclement for operations, in which case, irrespective of
the status shown by the weather sensors, the stations do not open up,
but instead stay in daysleep or wthsleep state.  This manual bad
weather status can be issued when the user has advance information on
e.g.~a weather condition approaching the site (such as a storm system). 
We can query the detailed status of the observer, including information
on the tasks scheduled, and the status of the individual hardware
components.

The node computer on each site hosts a web-page that graphically
displays the current and archival weather information in an
easy-to-read format.  The web-page also hosts the current and archival
images from both the web camera that monitors the \hsfour\ units and
the all-sky camera.  Nightly movies are created from both these cameras
and archived on the website.  These webpages allow HATSouth team
members across four continents to continuously check on both the
weather conditions and operations at all three sites, and to make good
decisions via manual intervention when conditions require.

\subsection{Observing and operation statistics}
\label{sec:stat}

Careful monitoring of the operational statistics is crucial for any
automated survey.  Such statistics can help in identifying and
rectifying areas of the operation that are responsible for lost
observing time.  They can also reveal failing hardware components, and
help in planning servicing missions to the sites, developing remote
workaround solutions, or devising tests to clarify the nature of such
failures.  Surprisingly little is known of the operation statistics of
automated surveys, and the reasons for failures are generally not
documented.

\reffig{obsstat} shows the weather and observing statistics for the
three HATSouth sites.  The light gray columns in \reffig{obsstat} show
the number of ``useful'' dark hours per night as a function of time for
the past two years, where ``useful'' is defined to be when weather
conditions meet the five criteria set out in \refsec{sites}.

The number of hours the CCDs spent exposing on-sky per night are
exhibited as dark gray bars in \reffig{obsstat}.  If we had a 100\%
duty cycle, our total exposure time spent on-sky for science frames
would be identical to the useful dark hours, i.e.~the dark gray bars
would perfectly match the light gray bars in \reffig{obsstat}.  In
reality, however, we lose time due to readout (25\,s per frame),
astrometry (12\,s), and re-pointing of the telescope (2\,s).  Also, we
take a bias frame after every 7 frames, and a 30\,s focus frame after
every 3 frames.  Altogether, under ideal conditions, our duty cycle is
73\%.  We plot the actual duty cycle in \reffig{obsstat} per night as a
function of time for all three sites (the bottom parts of each panel,
corresponding to the ratio of the dark gray to the light gray boxes). 
Chile, for example, is performing very stably at the 73\% level. 
Higher duty cycle values were present in earlier stages of the project,
when real-time astrometry and auto-focusing were not fully implemented. 
Consequently, the data quality during this ``shake-down'' period has
been of lower quality.

The major reason for lower duty cycles has been due to ice crystals
forming inside the chambers of many of the CCD cameras, which
necessitated a lengthy process of returning them to the manufacturer
for repair.  Curiously, this only affected CCD cameras at the HESS
(2011 April--July) and the SSO (e.g.~2010 March--July, and 2011
July--October) stations.  Another serious cause of operational downtime
was due to electronics being damaged by lightning storms both in
Namibia and Australia, leading to about 1 month of time lost for each
site.  Other events that caused more limited downtime include: Internet
outages (Namibia and Chile), instrument control software bugs, power
outages (instruments shut down and not recovered, but the weather
sensors operational), bush fire and excessive smoke (Namibia,
Australia), cables chewed by rodents (Chile), and damaged printer port
instrument control cards due to over-voltage.  Also there is downtime
during service missions,
as we upgrade and test equipment.

In spite of the various failures presented above, the overall duty
cycle of HATSouth has been very high.  As of 2011 December 31, the
\hsfour\ units opened up an average of 532 nights.  We gathered
$1\,060\,000$ \ccdsize{4K} images at 4\,minute integration time. 
Approximately 18 $8.2\arcdeg\times8.2\arcdeg$ fields on the sky have
been observed for an extensive time, accumulating more than $10\,000$
observations per field.  Importantly we have never had an operational
failure that resulted in a dome opening or remaining open in bad
weather conditions which would compromise the safety of the hardware. 
We therefore believe the HATSouth operations strike the right balance
between maximizing our duty cycle and minimizing the risk to the
hardware.

\section{Data flow and analysis}
\label{sec:dr}

\subsection{Data flow and data management}
\label{sec:dflow}

Many procedures relating to data have been carried over from the HATNet
project, and have been described previously by \citet{bakos:2004} or
\citet{bakos:2010}.  The most significant changes in the data of
HATSouth with respect to HATNet are: i) a much larger overall data
volume, ii) a more limited network bandwidth to the sites, iii) a more
complex data structure due to the larger number of hardware components.

The HATSouth network creates a significant volume of data.  Taking into
account the fraction of useful hours for each site, HATSouth (all six
\hsfour\ units) produces an average of 50\,GB/day raw compressed data
in the form of \ccdsize{4K} FITS files.  This corresponds to a rate of
$\sim150$\,GB/day of calibrated, uncompressed (real) science frames
from the network.  The average yearly yield of the HATSouth network is
19\,Terabytes (TB) of raw compressed pixel data, and 54\,TB of
calibrated science pixel data, plus $\sim6.5$\,TB of subsequent data
products, such as photometry files and \lcs.

On average HATSouth produces $\sim20$\,GB/day and 17\,GB/day of raw,
compressed data from our LCO and HESS sites respectively.  Transferring
these data to the HSDC in Princeton over the current Internet
infrastructure is not feasible.  We have therefore developed a scheme
whereby each month the server computer writes this data to Ultrium-4
800\,GB tapes.  The tapes are then shipped to the HSDC in Princeton. 
Local help is needed to change tapes and ship them to the HSDC.  This
is, in fact, the only routine manual interaction needed to run the LCO
and HESS sites.  The situation is slightly different for SSO, where a
gigabit data link between SSO and ANU in Canberra allows us to
relatively simply transfer all raw, compressed data from SSO to the
HSDC (averaging $\sim13$\,GB/day).  We still write data to Ultrium-4
800\,GB tapes at SSO for archival purposes, but regular shipping of
these tapes is not required.

\begin{figure*}[!ht]
\plottwo{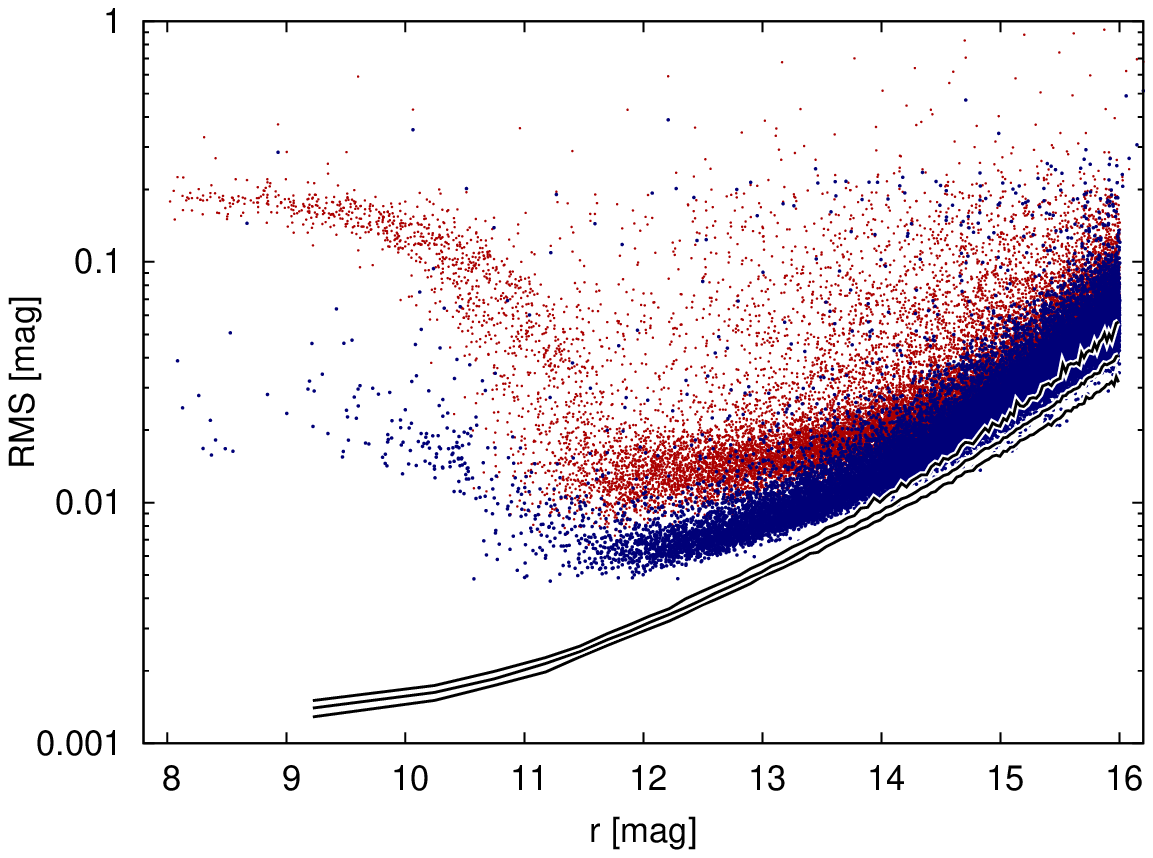}{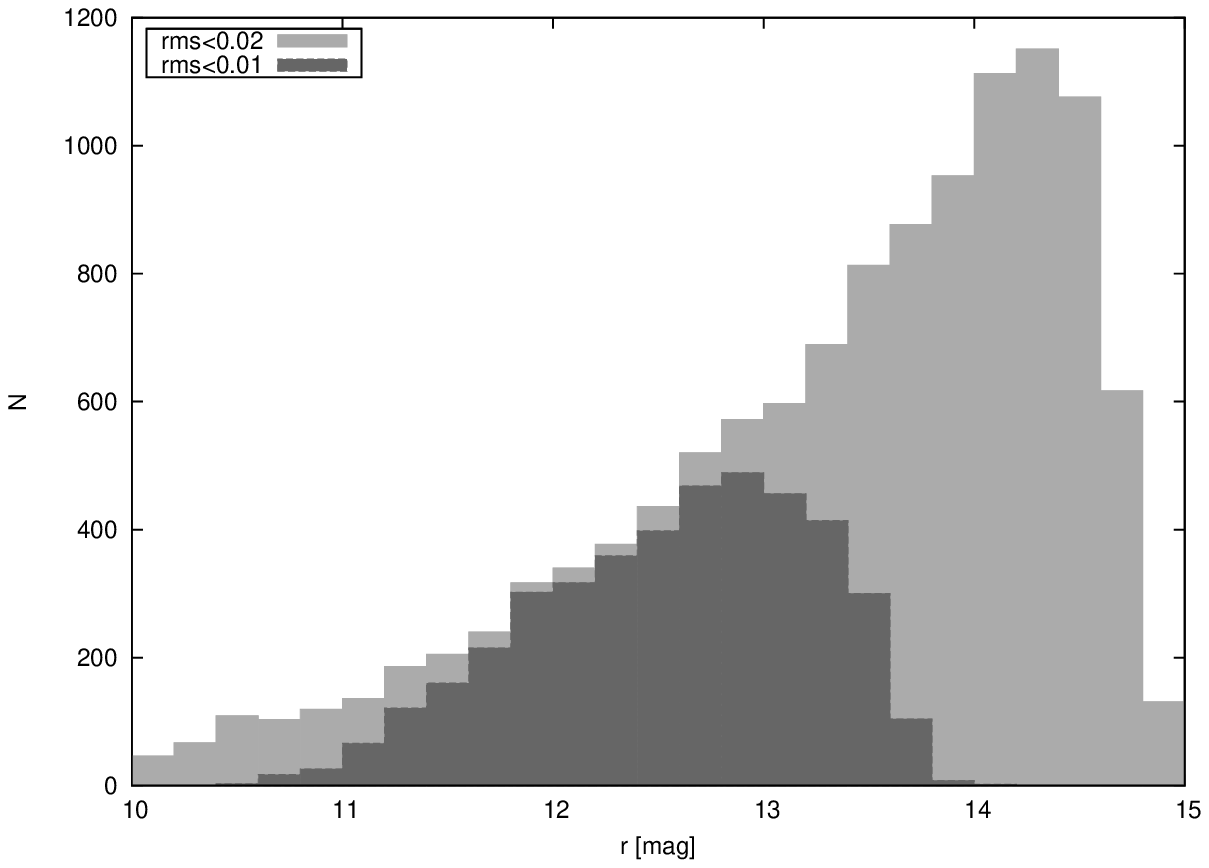}
\caption{
	The left panel shows the standard deviation of \lcs\ around the
	median as a function of Sloan $r$ magnitude.  The light (red)
	points refer to the fitted magnitudes ({\tt fitmag} in
	\refsec{dr}), and dark (blue) points refer to the TFA corrected
	{\tt tfamag} values.  Each point represents a \lc.  The standard
	deviation has been corrected for the effect of using a number of
	template stars in the TFA fitting procedure.  We use the optimal
	aperture (among three discrete choices) for each \lc, which choice
	is a function of the average magnitude of the star.
	We overplot the expected r.m.s.~for each \lc, taking into account
	the expected photometric error for each observation within the \lc\
	(calculated from the actual flux and sky background), and with
	scintillation noise \citep{dravins:1998} added in quadrature to the
	noise term.  We show the 10\% (lowest), 50\% (middle) and 90\%
	percentile, as a function of magnitude, from the distribution of
	the expected r.m.s.\@ The right panel shows the histogram of the
	number of stars in this particular (sparse) field with TFA
	r.m.s.~smaller than 0.01 as a function of \band{r} magnitude (dark
	gray) and similar data for TFA r.m.s~smaller than 0.02 (light
	gray).
\label{fig:rms}}
\end{figure*}

Data management for HATSouth is non-trivial, as we are dealing with a
large number of individual hardware components: 24 CCDs, 24 optical
tubes, 6 \hsfour\ telescope mounts at 3 sites.  Significant bookkeeping
is required for distinguishing between these in the calibrations and
subsequent data processing.  For example, optimal processing of a \lc\
requires the knowledge of which hardware combination produced various
subsets of the \lc.  We also need to keep track of the time evolution
of the components.  Occasional servicing or fine-tuning of the system
necessitates using a strict version control of the hardware.  We thus
maintain an extensive MySQL database describing all individual hardware
components, along with their time evolution, expressed through version
numbers.  We illustrate this process by way of four examples.
\begin{enumerate}
\item If a CCD is returned for servicing, the gain, readout noise, bad
pixel structure, cooling properties, and dark current structure may
change, and calibrating pre-servicing data with post-servicing
calibration frames (or the other way around) would be suboptimal.  In
such a case we thus introduce a new version number for the CCD, master
bias, and master dark.  Calibration is only permitted if version
numbers match.
\item If a CCD is removed from the OTA, the filter is cleaned, and the
CCD is installed back on the telescope, then the flatfield and pointing
version numbers are incremented.
\item If a Takahashi optical tube requires re-alignment of the optical
components, then the pointing model, flatfield version, and the PSF of
the system will change, and a new focus series needs to be acquired.
\item If the telescope mount home proximity sensor is replaced, the
pointing model of the system is incremented.  \end{enumerate}
There is a long list of other possible hardware changes, ranging from
the subtle (e.g.~cleaning the telescope front glass) to the extreme
(e.g.~replacing CCD cameras).  As long as the person servicing the
telescope notes the changes in the central database, the pipeline will
take them into account.

\subsection{Data processing}
\label{sec:dprocess}

Our data processing pipeline takes into account all of the data flow
and data management considerations set out in \refsec{dflow} to ensure
an optimal calibration procedure.  The pipeline has been developed in
the python programming language.  Routines where speed is an issue are
written in C or C++.  The pipeline is primarily run at the HSDC in
Princeton.  However the same pipeline, albeit with different sub-tasks,
is also run on the on-site control and server computers.

Each control computer writes all FITS frames from its \hsfour\ unit's
CCD cameras to a directory created for that given night.  Upon
successful closing of the actual observing session by the \vo\
(typically in the late morning, local time), that night's image
directory is ``released'', and the calibration pipeline on the control
computer is launched.  It generates master bias, dark and flatfield
frames for each CCD on the relevant \hsfour\ unit by combining the
individual dark, bias and flatfield frames taken over the course of the
night (average $\sim 30$ of each per night).  Bias and dark master
frames are the mean of the individual frames, after overscan correction
and outlier pixel rejection.  The master flatfield is a median average
of the individual frames, after overscan, bias and dark corrections,
and scaling the fluxes to the same median value in the center of the
frame.  Following this, all frames are transferred to our site server
computer through the local network, thus providing an immediate backup
of the data.  The raw frames on the control computer are then
compressed using a modified version of {\tt cfitsio/fpack}, and are
kept on the harddrives until they appear on our servers at the HSDC.

The site server computer produces a large variety of diagnostic plots
for each \hsfour\ unit.  Primarily these are plots of various telescope
and image properties as a function of time or hour-angle.  They are
grouped by the image type (bias, dark, flat, object) or observed field. 
Examples of parameters plotted are pointing errors, focus positions,
and the stellar profile.  We also generate very small size, compressed
(jpeg format) snapshots of selected regions of the science frames that
can be readily transferred over the Internet for review.  For certain
diagnostics we need to perform on-the-fly calibrations using the
on-site calibration master frames, however these on-site calibrated
frames are deleted after the diagnostics have been determined.  For all
sites the master calibration frames, diagnostic plots, and observing
logs (including MySQL database tables) are transferred to the HSDC each
day, typically equating to $<500$\,Mb of data transfer per site per
day.

Due to the requirement to ship tapes from the LCO and HESS sites to the
HSDC (see \refsec{dflow}), data arrives to the HSDC servers a few
months after acquisition.  Object frames are calibrated using the
already transferred master calibration frames.  Frames that have been
successfully read-in from tape to our HSDC servers, and which have the
same checksum as the original frames at the remote sites (checked via
{\tt md5sum}) are then purged from the remote site server and control
computers.  During the data transfer, we make sure that at least two
copies of the data are kept at any given time (including data on the
tapes).

The rest of the data processing takes place on the HSDC computer
cluster at Princeton University.  While ``quick-astrometry'' was
already run on the frames during observations (\refsec{oper}), we
refine the astrometric solutions using more time intensive procedures. 
Should the astrometry fail to find a solution for a frame, we exploit
the astrometric information from the neighboring CCDs (with known
geometrical offsets).  The astrometry uses sixth order polynomial fits
between pixel coordinates and the 2MASS catalog.  We then perform
aperture photometry at the fixed positions of the 2MASS stars in a
series of apertures.  Aperture photometry in 3 fixed radius apertures
is performed with tools originally developed as part of the HATNet
project \citep{pal:2009,bakos:2010}, and modified for our purposes. 
Photometry and subsequent data products are kept in binary format to
minimize storage requirements and improve performance in file I/O
operations.

Basic variations in the photometry due to extinction, changing PSF size
in the fixed aperture, etc., are removed by calibrating the photometry
of individual frames to the photometry of a reference frame.  This
reference is selected to be a frame taken under dark and transparent
sky conditions, and having the median FWHM of the other images.  The
procedure is done in an iterative way, excluding stars which display
significant brightness variations after applying the magnitude
transformation, as determined from a large ensemble of stars.  Such
variation might indicate the star is undergoing actual astrophysical
variability.  Stars are weighted by their individual formal photometric
errors in this fitting procedure.  After performing this correction to
all observations of a given field, a new reference frame is generated
by combining all corrected magnitudes, and the process is repeated by
using the r.m.s.~of their preliminary \lcs\ as weights.

\subsection{Light curves}
\label{sec:lcs}

We start with $\ordo(10^4)$ magnitude-fitted photometry files (one for
each frame) for a given field and CCD.  Each photometry file has
photometry information for each of the $\ordo(10^4)$ sources in the
frame.  The photometry files are then transformed into \lcs\
(information per source with $\ordo(10^4)$ observations), while
retaining the binary format.  The fitted magnitude \lcs\ (called {\tt
fitmag} \lcs) have remaining trends removed by decorrelating against
external parameters (resulting in {\tt epdmag} \lcs)
\citep[][]{bakos:2010} and then using the Trend Filtering Algorithm
(yielding {\tt tfamag} \lcs) \citep[TFA;][]{kovacs:2005}.

\begin{figure*}[!ht]
\plotone{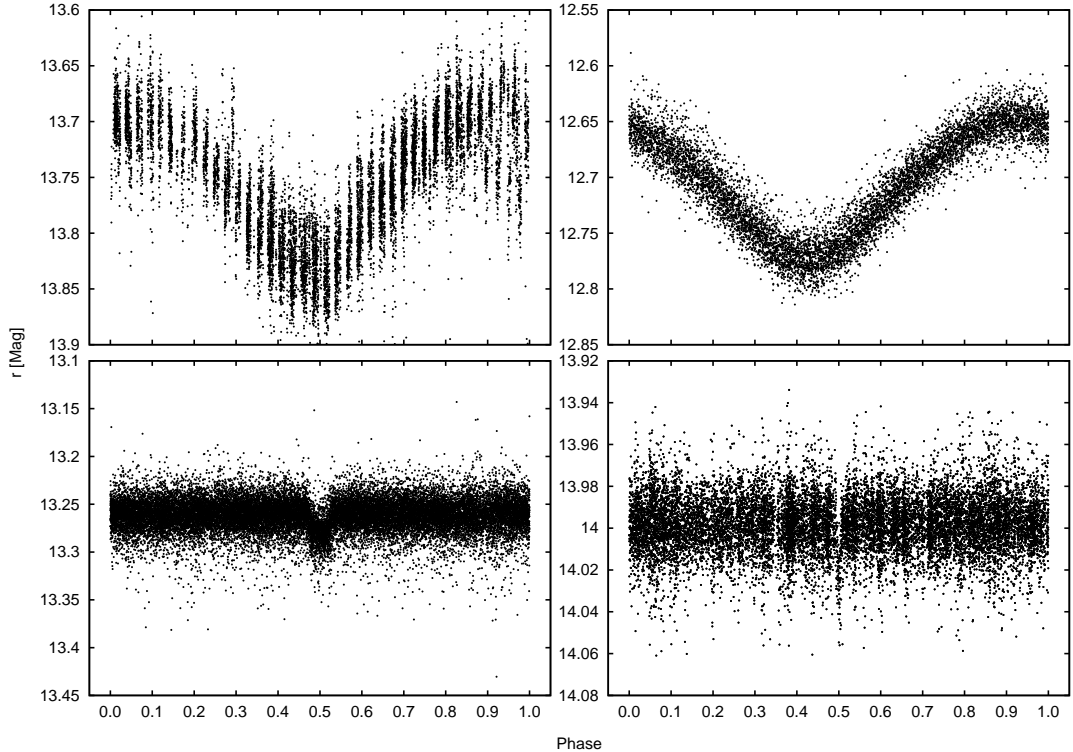}
\caption{
	Phased HATSouth light curves for four representative variable stars
	identified from the observations obtained to date.  These include a
	$P=37.9$\,d periodic variable (upper left), a $P=0.24$\,d periodic
	variable (upper right), a $P=1.4$\,d transiting hot Jupiter (lower
	left), and a $P=16.6$\,d transit candidate.  The upper two examples
	are {\tt epdmag} light curves (see \refsec{lcs}) with arbitrary
	reference phases, the lower two are {\tt tfamag} light curves and
	are displayed with the transits at phase 0.5.
\label{fig:examplelcs}}
\end{figure*}

For the brightest non-saturated stars, the resulting \lcs\ typically
reach a per-point photometric precision of $\sim 6$\,mmag r.m.s.~around
the median at 4-minute cadence.  These \lcs\ combine data from three
different \hsfour\ units, one each at LCO, the HESS site, and SSO.\@
The distribution of r.m.s.~values is shown in \reffig{rms} (left
panel), where red (light) points show the r.m.s.~of the {\tt fitmag}
\lcs, and black (dark) points exhibit the r.m.s.~of the final {\tt
tfamag} \lcs.  Each of the $\sim27000$ points represent a \lc\ with
$\sim8000$ observations, taken from a moderately sparse field at
galactic latitude $-40\arcdeg$.  Observations were contributed by all
three HATSouth sites.  The standard deviation has been ``unbiased'' so
as to correct for the effect of fitting the \lc\ with $\ordo(10^2)$
coefficients (the number of template stars) in the TFA procedure. 
Clearly, the detrending procedures significantly decrease the
r.m.s.~values of the \lcs.  Theoretical estimates have also been
overlaid in \reffig{rms} as solid lines.  In the left panel we show the
90th, 50th and 10th percentile curves for the expected r.m.s., as
calculated from the photon and sky background noise values for each
individual measurement, and an overall scintillation noise
\citep{dravins:1998} added in quadrature.  That is, we expect 90\% of
the r.m.s.~values would lie under the solid line on the top with these
noise sources included in the overall noise budget.  We are working on
clarifying and trying to rectify the discrepancy between the expected
and observed scatters of the \lcs.  There is room for improvement, and
we believe that a major contribution to this discrepancy may be due to
the complex sub-pixel structure of the microlensed and dual-gate
front-illuminated pixels (\refsec{ccd}), coupled with the anti-blooming
effect, and the relatively narrow PSF.  These effects appear to be not
very well corrected by the EPD and TFA procedures.  The right panel of
\reffig{rms} shows a histogram of stars in 0.2\,mag bins as a function
of \band{r} magnitude for stars having r.m.s.$<0.01$ and $<0.02$,
respectively.  The sub-percent r.m.s.~actually attained for stars with
$r<13.25$ allows us to achieve the primary science goal of HATSouth,
namely the discovery of transiting hot Jupiters, and even Neptune-size
planets (given the large number of data-points per field, and the often
uninterrupted time-series).  \reffigl{examplelcs} shows light curves
for four representative variable stars identified from the HATSouth
\lcs.

\subsection{Transit search}
\label{sec:tr}

We search the {\tt tfamag} \lcs\ for periodic transit events using the
Box-fitting Least Squares algorithm \citep[BLS;][]{kovacs:2002}.  We
then subject potential transit candidates to a number of automatic
filters to select reliable detections which are consistent with a
transiting planet-size object, and are not obviously eclipsing binary
star systems or other types of variables.  For example, we constrain
the maximum depth of the transit ($\lesssim 0.1$\,mag), we require at
least two transit events with altogether at least $50$ data-points in
transit, we limit the maximum radius of the transiting object to be
$\lesssim 2$\,\rjup\ (based on the transit depth, and assuming a
zero-age main sequence host star with stellar radius estimated from the
$J-K$ color index).  We also check for the difference in $\chi^2$
between the best fit transit model, and the best fit model where even
and odd transits are allowed to have different depths.  The maximum gap
in the phased \lc\ is also limited to be $\sim 0.2$.  We note that the
exact selection criteria may depend on the field, and we may perform
multiple runs of automatic selections with different criteria.
Altogether, the result is a machine-generated list of candidate TEP
hosts, including relevant light curve details (period, apparent transit
depth, etc.).  The automated transit candidate selection procedure
provides a manageable list of potential candidates which must then be
inspected by eye to select and prioritize the most promising targets
for follow-up.  Typically a few hundred to one thousand potential
candidates per $8.2\arcdeg\times8.2\arcdeg$ field are identified by the
automated procedures, which are then winnowed by hand to several dozen
candidates per field deemed worthy of follow-up.  First, the properties
of the \lc\ are analyzed, such as its BLS frequency spectrum, alternate
folding periods, out-of-transit variations, \lcs\ of neighboring stars,
appearance after running TFA with alternate templates, etc.  In
addition, a variety of resources are consulted, including archival
plates, 2MASS digital archives, appearance on HATSouth frames, proper
motions, and catalogs available through CDS/Vizier.  At the end of this
phase, relative priorities are assigned to surviving candidates and
appropriate facilities for follow-up are identified.

\section{Expected performance}
\label{sec:perf}

\begin{figure*}[!ht]
\plottwo{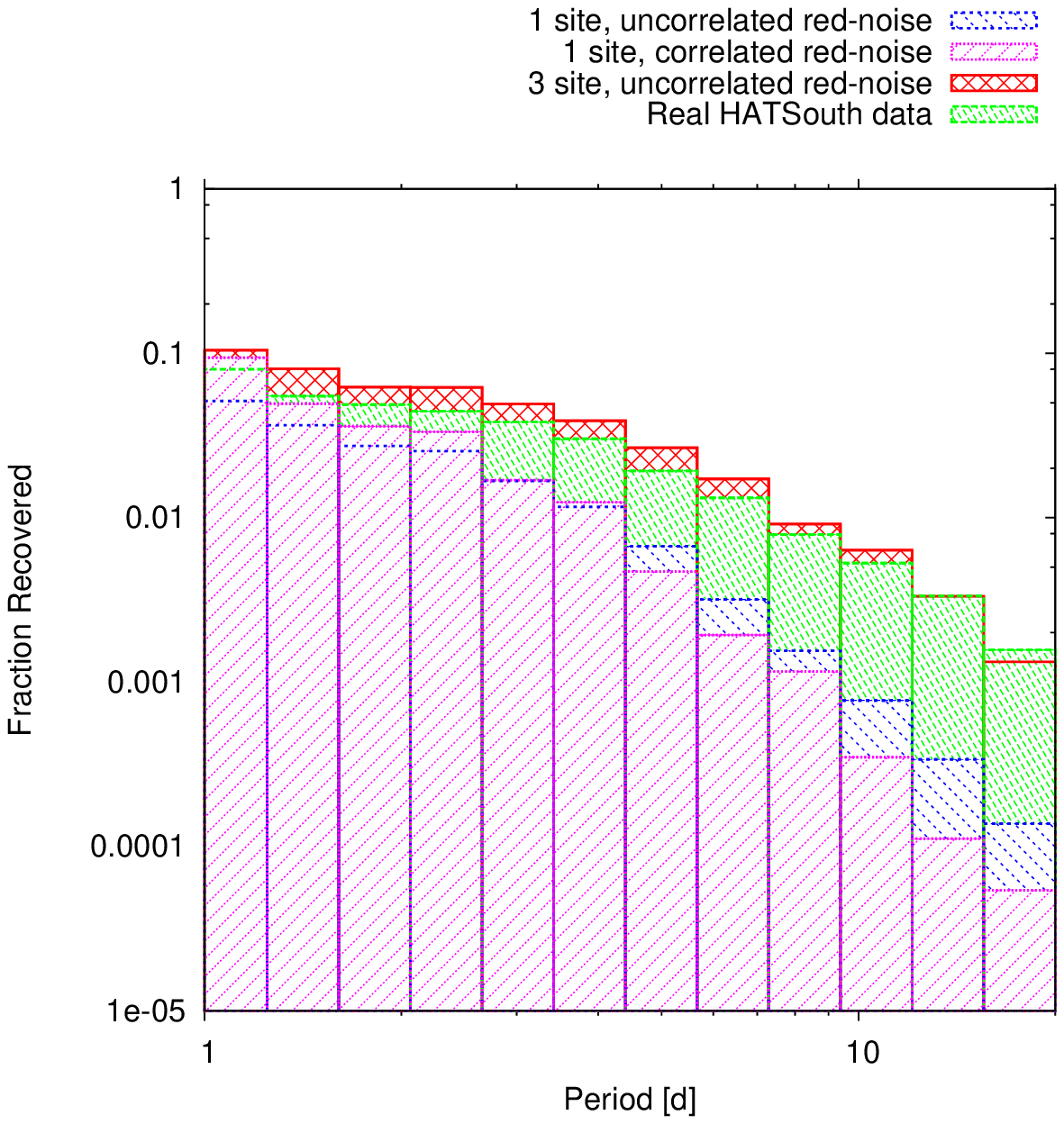}{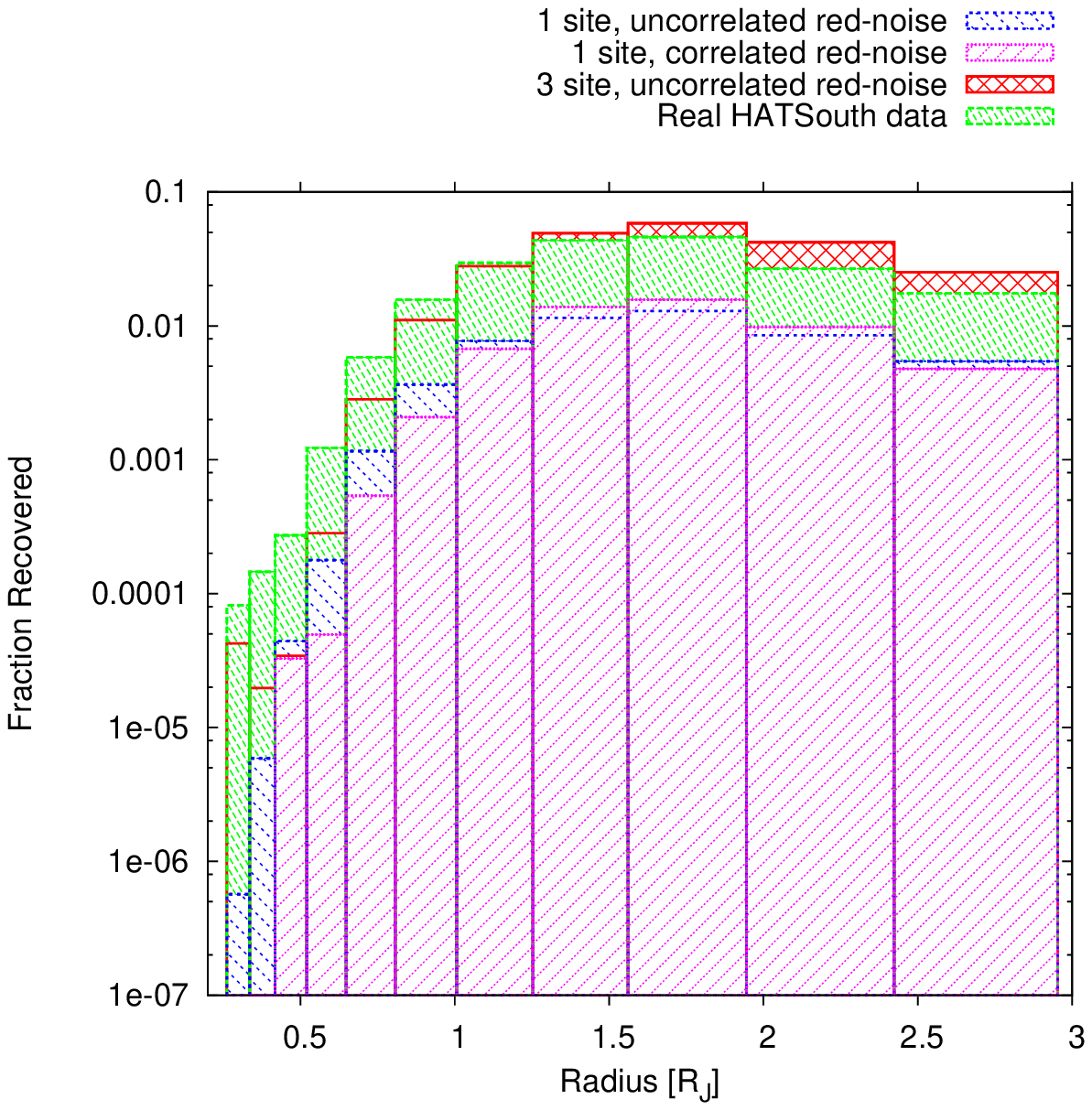}
\plottwo{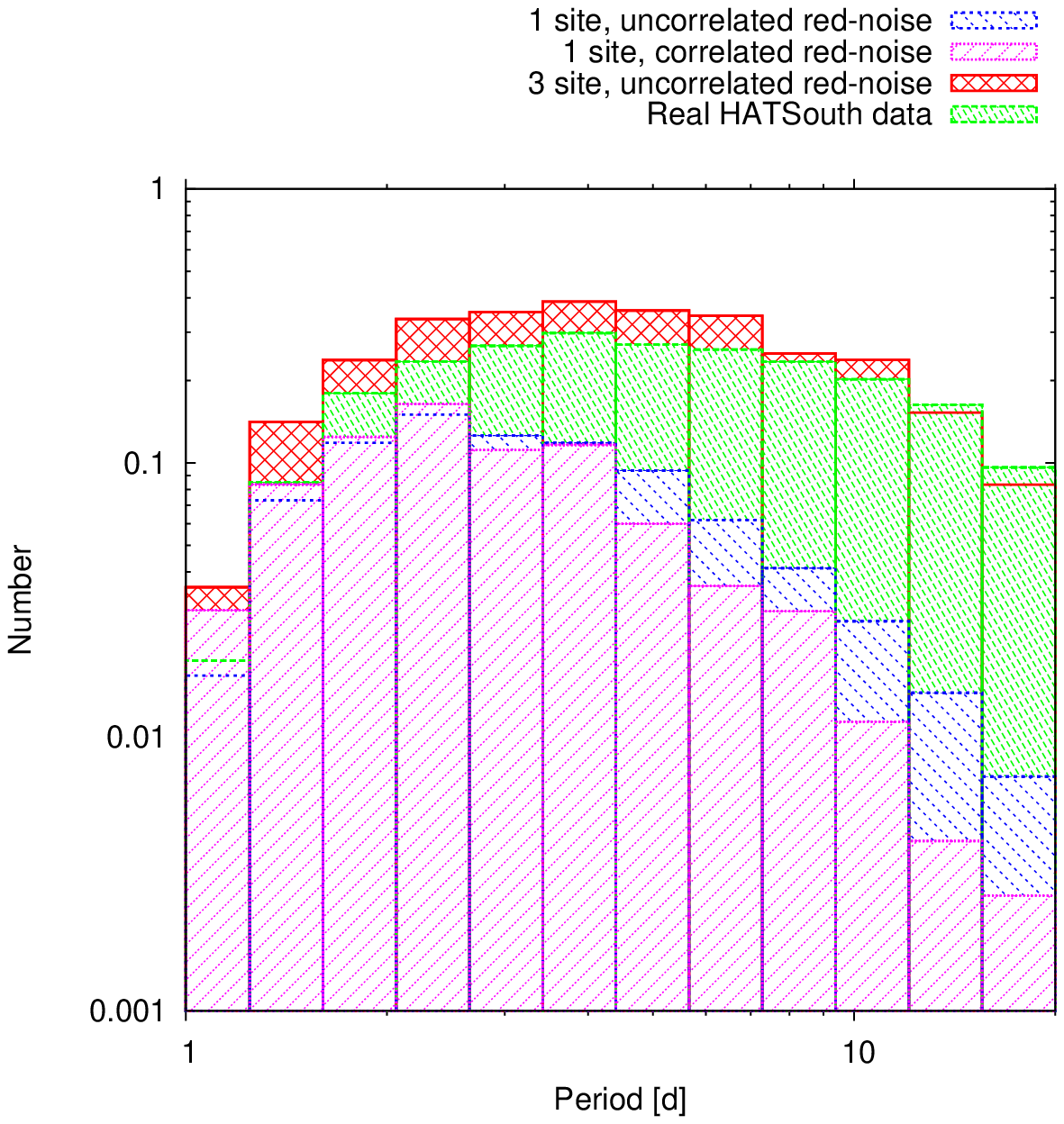}{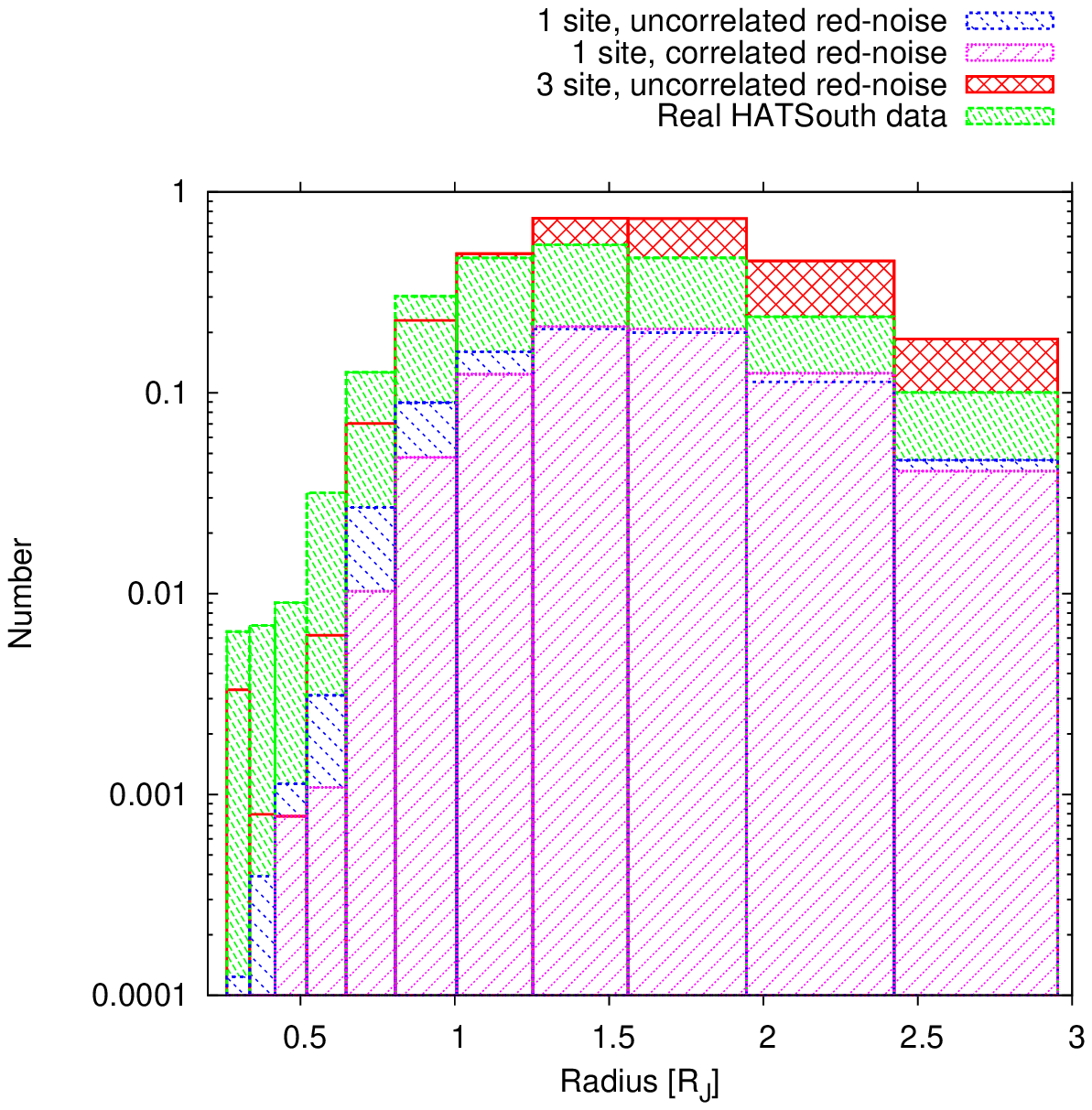}
\caption[]{
   The expected recovery rate of transiting planets as a function of
   period (upper left) and radius (upper right), marginalized in each
   case over the other parameter.  Note the logarithmic scales.  The
   recovery rates are displayed for the four sets of transit
   injection/recovery simulations listed in \refsecl{perf} (see the
   figure key; they are in the order of scenarios 1 through 4).  These
   are combined with the intrinsic planet period and radius
   distributions, and corrected for the geometric transit probability,
   to determine the expected planet yield for a single HATSouth field
   as described in \refsecl{perf}.  We show the yield as a function of
   period (lower left) and radius (lower right).
\label{fig:yieldsims}}
\end{figure*}

In order to estimate the expected yield of transiting planets from the
HATSouth survey we conduct transit injection and recovery simulations,
following the procedure summarized below (see also
\citealp{burke:2006}; \citealp{hartman:2009a}; and
\citealp{bayliss:2011}).  Additional details are provided in
Appendix~\ref{sec:transitrecoveryappendix}.
\begin{enumerate}
\item Use the Besan\c{c}on Galactic model \citep{robin:2003} to
  simulate a sample of stars which we might expect to observe in a
  typical HATSouth field.
\item Use the results from \cite{howard:2011}, based on the {\em
  Kepler} mission, to determine the underlying joint distribution of
  planet periods and radii over the ranges of HATSouth sensitivity. 
  This is a fundamental distinction from most previous attempts to
  estimate the expected planet yield of a transit survey where the
  planet distribution was either taken to be unknown, or was assumed to
  be constant over the period and radius ranges surveyed.
\item Generate a time-base (i.e.~a set of observation times to use for
  the light curve simulations) resulting from a two-month observing
  campaign.  This is generated taking into account the field visibility
  from each site, the exposure times and overheads, and the empirical
  weather statistics for each site.
\item Draw $> 5 \times 10^{5}$ samples from the above stellar and
  planetary distributions.
\item For each sample simulate a light curve with an injected
  transit. The light curve is generated with white and red noise
  appropriate for the magnitude of the star, with quadratic limb
  darkening parameters from \cite{claret:2004} appropriate for the
  stellar atmospheric parameters, random transit phase, and with $\cos
  i$ uniformly distributed between $0$ and $(R_{\star}+R_{P})/a$
  (i.e.~we assume a uniform distribution of orientations for the
  orbits, with the condition that transits must occur).  We assume
  circular orbits.
\item Apply the BLS and candidate selection algorithms used in the
  HATSouth search \refsec{lcs} to the simulated light curves to recover
  the transits.
\item Based on the recovery rate, the geometric transit probability
  for each simulation, and the total fraction of stars with planets
  from {\em Kepler} within the period and radius ranges used for our
  simulations, determine the total expected planet yield as a function
  of planet period and radius.
\end{enumerate}

\noindent 
We execute the above procedure four separate times for the following
scenarios:
\begin{enumerate}
\item Three imaginary HATSouth instruments at a single site 
  (we choose LCO as the site with the optimal weather statistics), red
  noise is uncorrelated between the instruments.
\item Three HATSouth instruments at a single site, red noise is
  correlated between the instruments.
\item One HATSouth instrument at each of the three sites.  This
	simulates the actual configuration of \hsfour\ units (even though
	the real setup has two units per site), using a generated time-base
	and simulated \lcs\ with appropriate noise parameters.
\item Transits are injected into {\em actual} HATSouth light curves for
  a field observed by 1 HATSouth instrument at each of the three sites
  over a two-month time-span.
\end{enumerate}

The results from these simulations are shown in \reffigl{yieldsims}. 
We plot the transit recovery rate both as a function of period and of
planet radius (i.e.~the fraction of transiting planets with a given
period or radius that would be recovered by HATSouth), and the expected
planet yield for a single HATSouth field (that is, the above, weighted
by the actual occurrence rate of planets).  The recovery rates and
yields are shown for each of the four scenarios listed above.  We find
that for a {\em single} HATSouth field observed over two months the
total expected planet yield is $0.85$, $0.77$, $2.9$ or $2.3$ planets
for scenarios 1 through 4, respectively (by total we mean integrated
over all radii and all periods).  These results clearly show the
significant increase in the expected planet yield by using a global
network (simulations 3 and 4); a three site global network has an
expected planet yield that is $>\!3$ times larger than the simulations
where all of the instruments are kept at a single site (simulations 1
and 2).  Assuming 12 fields observed per year, we expect to find $\sim
30$ transiting planets per year with HATSouth, including $\sim 1$
planet per year with $R\!<0.7\,R_{J}$ and $\sim 6$ planets per year
with $P > 10$\,d.  These numbers could be significantly improved by
reducing the instrumental red noise.  Of course, the final number of
planets heavily depends on the follow-up time invested in confirming
these candidates; this estimate assumes all candidates are followed up.

The expected $\sim 1$ planet per year with $R\!<0.7\,R_{J}$ could be a
Neptune-mass, or even a super-Earth mass planet.  It is worth noting
that the unique sensitivity (among ground-based transit searches) to
periods longer than 15 days opens up the possibility of finding
transiting super-Earths with orbits within the Habitable Zone
surrounding a mid-M dwarf.  Thus, a super-Earth orbiting an M5 dwarf
($\mstar\sim 0.21\,\msun$; $\rstar\sim0.27\,\rsun$) at or slightly
greater than a distance of 0.07\,AU, which corresponds to the inner
edge of the star's Habitable Zone \citep{kasting:1993}, would have an
orbital period of about 15 days or a little longer, well within the
reach of the HATSouth survey (\reffig{yieldsims}).

\section{Conclusions}
\label{sec:conc}

\begin{figure*}[!ht]
\plotone{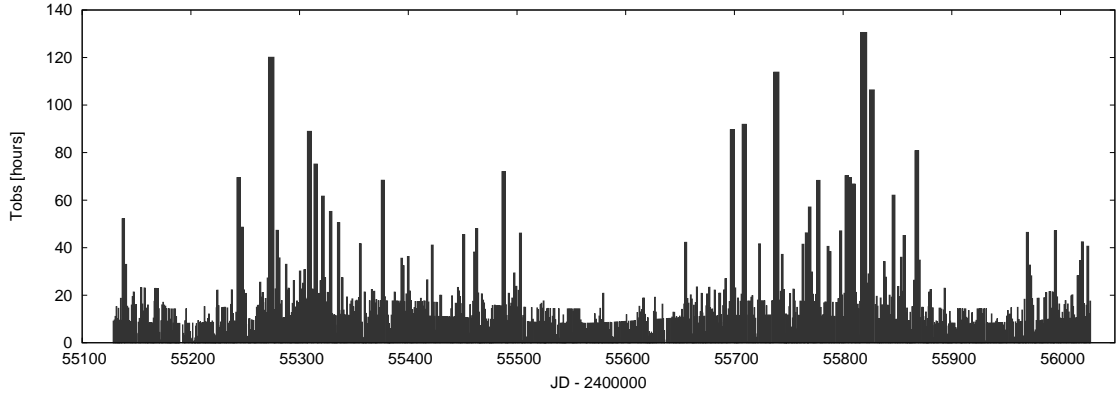}
\caption{
	Contiguous good weather stretches as a function of Julian Date. 
	The height (and width) of the boxes represent the number of dark
	hours that were clear from either of the LCO, HESS or SSO sites
	{\em without an interruption}.  Any bad weather longer than
	10\,minutes in duration is considered as an interruption.  The
	longest stretch of clear period was about 130\,hours (5.4\,days)
	long.
\label{fig:stretches}}
\end{figure*}

HATSouth is the world's first global network of identical and automated
telescopes capable of 24-hour observations all year around.  The
telescopes are placed at three southern hemisphere sites with
outstanding observing conditions (LCO in Chile, HESS site in Namibia,
SSO in Australia).  Long stretches of continuous observations are often
achieved.  \reffig{stretches} shows the contiguous blocks of clear
weather periods as a function of Julian Date for the past two years. 
The longest uninterrupted clear period, based on the detailed weather
logs for each of the three sites, is 130 hours long.  Relatively long
stretches, exceeding 24\,hours, are quite frequent.

HATSouth builds on the successful northern hemisphere HATNet project
\citep{bakos:2004}.  However, it implements numerous changes with
respect to HATNet.  Broadly speaking, we reach into a fainter stellar
population, having many more dwarf stars per square degree, thus
increasing our overall sample, and also having more dwarfs relative to
giant stars (which dilute the sample).  The fraction of K and M dwarfs
is also significantly higher, facilitating more efficient detection of
smaller planets, such as super-Earths.

Each of the three sites hosts two HATSouth instruments, called \hsfour\
units.  Each \hsfour\ unit holds four 0.18\,m, fast focal ratio
hyperbolic astrographs, tilted $\sim 4\arcdeg$\ with respect to each
other to produce a mosaic image spanning $8.2\arcdeg\times8.2\arcdeg$
on the sky, imaged onto four \ccdsize{4K} CCD cameras, at a resolution
of 3.7\pxs.  The photometric zero-point is $r\approx18.9$ (meaning
1\,ADU/s flux), and the 5-$\sigma$ detection threshold for the
routinely taken 240\,s images is $r\approx18.5$.  Stars become
saturated at $r\approx10.5\pm0.5$ in the 240\,s exposures, depending on
the focus (thus width of the stellar profile), and the degree of
vignetting at the position of the star.  We also monitor stars as
bright as $r\approx 8.25$ using shorter exposure (30\,s) images.

Meteorological conditions are monitored by a weather station (wind,
humidity, temperature, precipitation), a cloud detector (primarily
cloud cover), a lightning detector (forecasting lightning storms), and
an all-sky fisheye camera, all installed on our HATSouth control
building.  In addition, the individual \hsfour\ units are aided by a
hardwired rain detector and a photosensor (for avoiding daytime opening
of the dome).  Each \hsfour\ unit is controlled by a single
rack-mounted computer running Linux and Xenomai.  We have developed a
dedicated software environment for operating the telescopes, including
all hardware components, such as the dome, telescope mount and CCDs. 
Virtual Observer (\vo) is the intelligent software, managing all
aspects of running the observatory, such as preparing devices,
scheduling observations, monitoring the weather, handling exceptions,
communicating with the outside world, and logging events and
observations.

The network monitors selected fields on the southern sky for about
2\,months per field.  Fields are selected in a way that all dark time
during the night is used.  A significant effort is made to optimize the
data quality, by re-focusing the optics every $\sim15$\,minutes,
running real-time astrometry after the frames, and adjusting the
pointing of the mount.

This combination of precise weather monitoring, the use of a very
stable operating system, and running a dedicated software environment
has resulted in very robust operations.  Indeed, the HATSouth network
has taken over 1 million \ccdsize{4K} science frames during its initial
2 years of operations.  The six \hsfour\ units have each opened up an
average of $\sim500$ nights so far, without a single case of opening
when weather conditions were not suitable.

We have developed a scheme that reduces the amount of data transferred
from the remote sites to the HSDC in Princeton.  The reduction pipeline
keeps track of the large number of individual hardware components (24
OTAs, 24 CCDs, etc), and maintains a version control for the above,
since hardware may change due to routine maintenance or instrument
repairs.  The current reduction procedure, after application of the
Trend Filtering Algorithm \citep[TFA][]{kovacs:2005}, yields \lcs\
reaching 6\,mmag r.m.s.~at 240\,s cadence at around the saturation
limit.  The \lcs\ are searched for transit candidates using a
well-established methodology that has been developed for HATNet, and
relies on the BLS \citep[][]{kovacs:2002} algorithm and
post-processing.  The network is producing high quality planetary
transit candidates, and general variability data (\reffig{examplelcs}). 
Follow-up observations for these candidates are being performed as an
intensive team-effort, and will be described in subsequent
publications.

We have run detailed, realistic simulations on the expected yield of
transiting planets from the HATSouth network.  The simulations take
into account the noise characteristics of the instruments, the weather
pattern, the observing windows, the stellar population, the expected
planetary population based on recent {\em Kepler} results, and our
search methodology for transits.  We compared two basic scenarios: all
\hsfour\ telescopes are at a single site and observing the same field
at higher S/N, or the telescopes are spread out to the current setup,
and observe the same field at lower S/N (per unit time), but at much
higher fill-factor.  The simulations were performed both with
uncorrelated and correlated ``red''-noise components in the \lcs.  The
results clearly prefer the networked setup, predicting a three-fold
increase in the number of detected transiting planets, as compared to a
single site setup.  The long stretches of observations
(\reffig{stretches}), and uncorrelated noise between the stations, are
clearly fundamental in this increased yield.  Notably, the fraction of
planets recovered at $P\approx10$\,d period is about 10 times that of a
single-site installation (\reffig{yieldsims}, top left panel), and the
planetary {\em yield} after taking into account the {\em Kepler}
distribution of planets and the geometric transit probability is also
$\sim10$ times higher than for the single-site installation
(\reffig{yieldsims}, bottom left panel).  The peak sensitivity occurs
at $P\approx 6$\,days (marginalized over all planetary radii), and
exhibits much slower decline towards long periods than the single-site
setup.  Similarly, there is a significant increase (factors of 3 to 10)
in the detection efficiency as a function of planetary radius,
especially at small radii.  The peak sensitivity, however, occurs at
$\sim 1.5\,\rjup$ for both the networked and single-site setups. 
Altogether, we expect that HATSouth is capable of the detection of
$\sim30$ transiting extrasolar planets per year, pending follow-up
confirmation of these candidates.

Note that while the HATSouth stellar sample is somewhat fainter than
that of HATNet, the candidates are still within the reach of follow-up
resources available in the southern hemisphere.  Examples for such
spectroscopic follow-up resources (including those with reconnaissance
or high precision radial velocity capabilities) are the Wide Field
Spectrograph \citep[WiFeS;][]{dopita:2007} on the ANU~2.3\,m telescope,
FEROS on the MPG/ESO 2.2\,m telescope, CORALIE on the Euler 1.2\,m
telescope, HARPS on the ESO 3.6\,m telescope (all at La Silla, Chile),
the Echelle spectrograph on the 2.5\,m du Pont telescope at LCO, and
the UCLES spectrograph on the 3.9\,m AAT at SSO.

Global networks of telescopes present a powerful way of studying
time-variable astronomical phenomena.  By coupling this with telescopes
that are identical and fully automated, it is possible to undertake
large, long duration surveys that would have been completely unfeasible
with manually operated or single site facilities.  HATSouth is the
first of many projects that will utilize the combination of these two
concepts over the next decade and, we hope, make many exciting
discoveries in the process.

\acknowledgements 

\paragraph{Acknowledgements}
Hardware costs for the HATSouth project were funded by NSF MRI grant
NSF/AST-0723074.  Construction of the infrastructure at the HESS site
was funded by MPIA, and similar costs at SSO enjoyed partial funding
from the ANU.  We thank the Smithsonian Astrophysical Observatory for
the Internal Research and Development (IR\&D) funds that supported
visiting students G\'abor Kov\'acs and Zolt\'an Csubry during the
2008--2009 development phase of the project.  Following the
installation and calibration, the overall operation of the HATSouth
network has been supported by NASA grant NNX09AB29G.

We wish to thank directors Wendy Freedman (Carnegie -- Las Campanas
Observatory), Miguel Roth (Las Campanas), Harvey Butcher (ANU -- Siding
Springs Observatory), and Werner Hofmann (MPIK -- HESS site) for their
generous support towards accommodating the HATSouth instruments at the
respective sites.  Our home-base for development, procurement,
installation, and early operation of the HATSouth network was at the
Smithsonian Astrophysical Observatory.  We greatly appreciate the
outstanding support of SAO director Charles Alcock, and associate
directors Nancy Brickhouse and Roger Brissenden.  We also thank Matt
Holman for his help in the initial phase of the project.

Installation to the respective HATSouth sites required team-work in a
hectic schedule.  We thank F.~R\'ozsa for joining our installation team
to Namibia and Australia.  At Las Campanas, we are indebted to the
whole staff for their help in installation, and particularly so to
Francisco Figueroa, Marc Leroy, Juan Espoz, Juan Espoz, Jr., Andr\'es
Rivera, Patricio Pinto, and Henry Cortes for their very generous
support.  Don Hector and the kitchen staff at LCO provided delicious
meals that kept us going through the long installation days.  We also
thank Earl Harris (Carnegie) for his assistance in the transportation
of telescope components to the site.
At the HESS site, we are most grateful for the help of Toni Hanke, site
manager, technical support Albert Jahnke, and our hosts Adele and
Joachim Cranz.  We greatly benefited from discussions with Michael
Panther and German Herman (MPIK) in the initial planning phase of the
project.
At ANU's Siding Springs Observatory, we thank Geoff White for his help
during installation and the first two year's of operation.  Tammy
Roderick at ANU helped identify many issues in relation to the first
set of HATSouth images and we thank her very much for her contribution
to the project.  We greatly appreciate the help we received from Gabe
Bloxham at ANU in relation to the telescope optics.  We also thank
Colin Vest at ANU for his help in servicing the Australian telescopes. 
We appreciate Vince O'Connor's help in administrative matters regarding
the installation.
G.\'A.B.~was supported as an NSF Fellow (AST-0702843) by SAO during the
development of this project.  G.\'A.B.~is extremely thankful for the
family of P.~S\'ari for hosting and operating the prototype instrument
in their garden at P\'ecel, Hungary, for an extended period of almost
one year.  G.\'A.B.~also wishes to thank the SAO administration for the
swift management of the relevant grants, in particular to P.~Sozanski,
J.~Lendall, N.~Rathle, L.~Linarte and E.~Dotts.  We thank T.~Szklen\'ar
for his expert help in the system management of computers during 2011
at the remote sites and at Princeton.  We also thank Gy.~Medgyesi for
his help in the mechanical design of the HATSouth units.  We thank
P.~J\'on\'as and his family for their help in the Australian
installation.
A.J.\ acknowledges support from Fondecyt project 1095213, Ministry of
Economy ICM Nuclei P07-021-F and P10-022-F, Anillo ACT-086, BASAL CATA
PFB-06 and FONDAP CFA 15010003.  V.S.\ acknowledges support form BASAL
CATA PFB-06.  M.R.~has been supported by ALMA-CONICYT projects 31090015
and 31080021, and by a FONDECYT postdoctoral project 3120097.  G.K.~is
supported by RoPACS, a Marie Curie Initial Training Network funded by
the European Commission’s Seventh Framework Programme.

\clearpage
\input{biblio.tex}

\appendix

\section{Details of the Planet Yield Simulations}
\label{sec:transitrecoveryappendix}

The general procedure that we follow to determine the expected planet
yield with HATSouth is outlined in \refsecl{perf}, here we provide
additional details concerning this procedure.

We use the online Besan\c{c}on Galactic model simulator to generate a
sample of stars observed in a typical HATSouth field centered at
coordinates $\alpha = 300^{\circ}$, $\delta = -22.\!\!^{\circ}5$, which
is at $b = -24.\!\!^{\circ}6$ Galactic latitude.  Importantly, this
model gives both physical parameters (mass and radius) and observed
parameters (magnitudes in various passbands) for each star, which are
needed to simulate the transit light curves.

We use the underlying planet period and radius distribution from
\citet[][hereafter H11]{howard:2011}. This includes Equation~4 from
H11 for the planet radius distribution:
\begin{equation}
\frac{{\rm d}f(R)}{{\rm d}\log R} = k_{R} R^{\alpha}
\end{equation}
with $\alpha = -1.92 \pm 0.11$ and $k_{R} = 2.9^{+0.5}_{-0.4}$ for
planets with $P\!<\!50$\,d, and Equation~8 from H11 for the planet
period distribution:
\begin{equation}
\frac{{\rm d}f(P)}{{\rm d}\log P} = k_{P}P^{\beta}\left 
	(1 - e^{-(P/P_{0})^{\gamma}} \right )
\end{equation}
with values for $k_{P}$, $\beta$, $P_{0}$ and $\gamma$, which depend on
the planet radius, taken from Table~5 of H11.  They found that 16.6\%
of GK dwarf stars have planets in the range $P\!<\!50$\,d and
$2\,R_{\earth}\!<\!R\!<\!32\,R_{\earth}$.  We restrict our simulations to
the range $P\!<\!20$\,d and $3\,R_{\earth}\!<\!R\!<\!32\,R_{\earth}$. 
Integrating the H11 distribution over the restricted ranges yields a
planet occurrence frequency of 1.73\%.  This is used, together with the
geometric transit occurrence probability for each planet, the total
number of stars observed in the field, and the total number of fields
observed by HATSouth per year, to scale the recovery frequency from our
simulations to determine the expected yearly planet yield from
HATSouth.

In drawing samples of stars and associated planets we assume that the
stellar and planet distributions are independent.  This is known to be
a false assumption when considering stars over the spectral range from
M to A \citep[e.g.][]{johnson:2011}.  However, the distribution of
spectral types for the dwarf stars observed by HATSouth is expected to
be fairly similar to that observed by {\em Kepler}, with perhaps a
slight bias towards hotter stars in HATSouth given the brighter
magnitude limits of the survey.  Because the occurrence of massive
planets on close-in orbits increases for hotter stars, assuming the
{\em Kepler} planet distribution applies for all stars in the HATSouth
survey most likely underestimates the total number of planets to which
HATSouth is sensitive.

We follow two approaches to generating light curves with realistic
noise properties.  One approach is to simply use the real (observed)
light curves (simulation 4 in \refsecl{perf}) choosing for each
simulated star the observed star from a given field that is closest to
the simulated star in magnitude.  The other approach is to simulate a
light curve with both white and red noise (simulations 1 through 3 in
\refsecl{perf}).  To do this we follow the wavelet-based procedure
given in Section~4.2 of \cite{mccoy:1996} (see also
\citealp{carter:2009}) to generate a uniformly sampled time-series with
$\sim f^{-0.99}$ red noise using the publicly-available {\sc
  Vartools} program \citep{hartman:2008}. This time series is then
interpolated onto the time-base of the simulated observations, and
scaled to have a standard deviation of $5$\,mmag (our conservative
estimate of the red-noise in the HATSouth light curves). To this
light curve we add Gaussian random noise with standard deviation equal
to the expected light curve standard deviation, based on the
photometric errors (including photon noise from the star, sky noise,
and read-noise), of the observed star from a given field that is
closest to the simulated star in magnitude.

We use {\sc Vartools} to inject \cite{mandel:2002} model limb-darkened
transit light curves into the simulated light curves.  We assume only
one transiting planet per star.  Before injecting the transits we
dilute the model signal by a factor of 0.8 which, based on our
experience with HATNet, is the typical factor by which the TFA
detrending algorithm, as applied to the HATSouth light curves, reduces
the transit depths.  We then apply BLS at the fixed period of the
injected transit to determine the S/N of the transit in the light
curve.  In the great majority of cases the transits are much too
shallow relative to the light curve noise to have any hope of detecting
them (the stellar distribution increases towards fainter stars, while
the planet distribution increases towards smaller radius planets on
longer period orbits), we therefore use this quick cut on the S/N of
the injected transit to immediately reject undetectable transits
without executing the full BLS transit search, and reserving the
computationally expensive search only for those simulations for which
there is some chance that the transit could be recovered.

To simulate as closely as possible the HATSouth survey, as executed, we
apply the same BLS search, peak identification, and automated candidate
selection routines to the simulations passing the aforementioned S/N
cut as we apply in the actual survey.  The selection routines are
applied to the top five peaks in the BLS spectrum of a light curve.  To
consider a simulation to be recovered we further require that one of
the peaks that passes the automated selections corresponds to a
frequency that is within $0.02$\,d$^{-1}$ of the injected frequency. 
To account for the possibility that our by-eye selection of candidates
rejects real planets which pass the automatic selections, we inspected
by eye a randomly selected subset of the simulations which passed the
automatic cuts.  As a control, we mixed into this subset a random
sample of simulated light curves for which the injected transit was not
recovered, but for which a significant peak was identified in the BLS
spectrum.  We found that 95\% of the automatically recovered transits
also passed the by-eye selection (ranging from about 70\% of the
simulations with S/N near the cut-off value, to 100\% of the
simulations with S/N$ > $12.5), while 90\% of the non-recovered
transits were rejected by our by-eye selection.

The total expected yearly yield of planets with periods between $P_{\rm
min}\!<\!P\!<\!P_{\rm max}$ and radii between $R_{\rm min}\!<\!R\!<\!R_{\rm
max}$ is given by:
\begin{equation}
N(P,R) = \frac{N_{\rm obs}}{N_{\rm sim}}f_{\rm tot}\sum_{i=1}^{N_{\rm sim}}
	\theta(R_{i} - R_{\rm min})\theta(R_{\rm max} - R_{i})
	\theta(P_{i} - P_{\rm min})\theta(P_{\rm max}-P_{i})
	\delta_{i}f_{\rm eye}(SN_{i})\frac{R_{\star,i}+R_{i}}{a_{i}}
\end{equation}
where $N_{\rm obs}$ is the number of dwarf stars expected to be
surveyed per year (given by the number of dwarf stars in the
Besan\c{c}on simulation, times 12 fields observed per year), $N_{\rm
  sim}$ is the number of simulations conducted, $f_{\rm tot}$ is the
total fraction of stars with planets within the period and radius
ranges of the simulations (1.73\%), and the sum is over all
simulations, with $\theta(x) = 1$ for $x > 0$, $\theta(x) = 0$ for 
$x\!<\!0$, $\delta_{i} = 1$ if the simulated transit is recovered and $0$
if not, $f_{\rm eye}(SN_{i})$ is the fraction of automatically
selected candidates with transit S/N$\sim SN_{i}$ that pass the by-eye
selection, and $(R_{\star,i}+R_{i})/a_{i}$ is the geometric
probability of transit for simulation $i$.

\end{document}

%% file: newcommand_gen.tex
\newcommand{\ctbd}[1]{}

\newcommand{\lc}{light curve}
\newcommand{\lcs}{light curves}

\newcommand{\ccdsize}[1]{\ensuremath{\rm #1\times\rm#1}}

\newcommand{\band}[1]{\ensuremath{#1}~band}
\newcommand{\ordo}{\ensuremath{\mathcal{O}}}

\newcommand{\ms}{\ensuremath{\rm m\,s^{-1}}}

\newcommand{\C}{\ensuremath{\rm ^{\circ}C\;}}
\newcommand{\el}{\ensuremath{e^-}}
\newcommand{\mum}{\ensuremath{\mu m}}

\newcommand{\sqarcdeg}{\ensuremath{\Box^{\circ}}}
\newcommand{\pxs}{\ensuremath{\rm \arcsec pixel^{-1}}}

\newcommand{\eladu}{\ensuremath{\rm \lbrack \el/ADU \rbrack}}

\newcommand{\rsun}{\ensuremath{R_\sun}}
\newcommand{\msun}{\ensuremath{M_\sun}}

\newcommand{\rstar}{\ensuremath{R_\star}}
\newcommand{\mstar}{\ensuremath{M_\star}}

\newcommand{\rearth}{\ensuremath{R_\earth}}

\newcommand{\rjup}{\ensuremath{R_{\rm J}}}
\newcommand{\mjup}{\ensuremath{M_{\rm J}}}

\newcommand{\reffig}[1]{Fig.~\ref{fig:#1}}
\newcommand{\refsec}[1]{\mbox{\S\ \ref{sec:#1}}}

\newcommand{\reftab}[1]{Tab.~\ref{tab:#1}}

\newcommand{\reffigl}[1]{Figure~\ref{fig:#1}}
\newcommand{\refsecl}[1]{\mbox{Section \ref{sec:#1}}}

%% file: newcommand_spec.tex
\newcommand{\dome}{{\tt dome}}
\newcommand{\scope}{{\tt scope}}
\newcommand{\wthdaemon}{{\tt wthdaemon}}
\newcommand{\mountsrv}{{\tt mountsrv}}
\newcommand{\vo}{{\tt vo}}

\newcommand{\ccdsrv}{{\tt ccdsrv}}
\newcommand{\hatbb}{{\tt hatbb}}
\newcommand{\hsfour}{HS$_4$}

\newcommand{\bias}{{\tt bias}}
\newcommand{\dark}{{\tt dark}}
\newcommand{\skyflat}{{\tt skyflat}}
\newcommand{\monf}{{\tt monfield}}
